\documentclass[final,preprint,3p,square,times,twocolumn]{elsarticle}
\usepackage[german,american,english]{babel}
\usepackage{graphicx}
\usepackage{graphics}
\usepackage{dcolumn}
\usepackage{bm}
\usepackage{amssymb}
\usepackage{amsmath}
\usepackage{amsfonts}
\usepackage{epsfig}
\usepackage{mathbbol}
\usepackage{dcolumn}
\usepackage{subfigure}
\usepackage{hyperref}
\usepackage{dsfont}
\biboptions{sort&compress}
\hypersetup{
   colorlinks,
    citecolor=blue,
   filecolor=blue,
    linkcolor=blue,
    urlcolor=blue
}

\newcommand {\pd} [2] {\frac{\partial #1}{\partial #2}}

 \newcommand {\beq}{\begin{equation}}
\newcommand {\eeq}{\end{equation}}
\newcommand {\beqn}{\begin{eqnarray}}
\newcommand {\eeqn}{\end{eqnarray}}
\newcommand {\bit}{\begin{itemize}}
\newcommand {\eit}{\end{itemize}}

\newcommand{\ba}{\begin{array}{rl}}
\newcommand{\ea}{\end{array}}
\newcommand{\bc}{\begin{cases}}
\newcommand{\ec}{\end{cases}}

\newcommand{\dps}{\displaystyle}

\newcommand{\om}{\iffalse}
\journal{Physica E}
\begin{document}
\begin{frontmatter}

\title{Coulomb drag in topological insulator films}

\author{Hong Liu}
\address{ICQD, Hefei National Laboratory for Physical Sciences at the Microscale, University of Science and Technology of China, Hefei 230026, Anhui, China}
\author{Weizhe Liu}
\author{Dimitrie Culcer$^*$}
\address{School of Physics, The University of New South Wales, Sydney 2052, Australia}

\date{\today}

\begin{abstract}

We study Coulomb drag between the top and bottom surfaces of topological insulator films. We derive a kinetic equation for the thin-film spin density matrix containing the full spin structure of the two-layer system, and analyze the electron-electron interaction in detail in order to recover all terms responsible for Coulomb drag. Focusing on typical topological insulator systems, with film thicknesses $d$ up to 6\text{nm}, we obtain numerical and approximate analytical results for the drag resistivity $\rho_\text{D}$ and find that $\rho_\text{D}$ is proportional to $T^2d^{-4}n^{-3/2}_{\text{a}}n^{-3/2}_{\text{p}}$ at low temperature $T$ and low electron density $n_{\text{a,p}}$, with a denoting the active layer and p the passive layer. In addition, we compare $\rho_{\text{D}}$ with graphene, identifying qualitative and quantitative differences, and we discuss the multi valley case, ultra thin films and electron-hole layers.
\end{abstract}

\begin{keyword}
Electron-electron interactions \sep Coulomb drag \sep Topological insulator
 \end{keyword}
\end{frontmatter}

\section{Introduction}
Three-dimensional topological insulators (3DTIs) are a novel class of bulk insulating materials that possess conducting surface states with a chiral spin texture \citep{KaneMele_QSHE_PRL05,Hasan_TI_RMP10,Qi_TI_RMP_10,Moore_TRI_TI_Invariants_PRB07,Ando-TI, Culcer_TI_PhysE12, Tkachov_TI_Review_PSS13,Yong-qing_FOP_2012, Moore-nature}. Thanks to their topology, these surface states remain gapless in the presence of time-reversal invariant perturbations. Following their initial observation \citep{Urazhdin_PRB04, Hsieh_BiSb_QSHI_Nature08, Hsieh_BiSb_QmSpinTxtr_Science09, Xia_Bi2Se3_LargeGap_NP09, Hsieh_Bi2Te3_Sb2Te3_PRL09, Taskin_TI_eh_PRL11, Zhang_Bi2Se3_Film_Epitaxy_APL09, Wang_Bi2Te3_Ctrl_AM11, Ren-PRB-2010, Benjamin_nature_2011}, improvements in TI growth have made them suitable for fundamental research \citep{Dohun_Surface_nphy_2012, Fuhrer_nature_2013, Brune_prx_2014, Hellerstedt_APL_2014}. Although the reliable identification of the surface states in transport, which remains the key to TIs becoming technologically important, has remained elusive, a number of experiments have successfully identified surface transport signatures in isolated samples. These were initially mostly singled out via quantum oscillations or in gated thin films \citep{Ren-PRB-2010, Benjamin_nature_2011, Fuhrer_nature_2013, Brune_prx_2014, Dohun_Surface_nphy_2012}. Recently, four-point transport measurements on clean surfaces in an ultrahigh vacuum have reported a surface-dominated conductivity \citep{Lucas_nl_tran_2014}. A current induced spin polarization also constitutes a signature of surface transport \citep{Burkov_TI_SpinCharge_PRL10, Culcer_TI_Kineq_PRB10} and was reported in recent experimental studies \citep{LiC.H_spin_polarization_nature2014, Jianshi_nl_spin_polarized_2014, Kastl_surface_tran_nature_2015}. Magnetic TIs have also been successfully manufactured \citep{Hor_DopedTI_FM_PRB10,Jinsong-Zhang-Yayu-Wang-2013-science,Collins-McIntyre_Cr_Bi2Se3_2014}, and the anomalous \citep{Culcer_TI_AHE_PRB11} and quantum anomalous Hall effects \citep{Yu_TI_QuantAHE_Science10,LuZhao_TITF_transport_PRB2013} have been detected \citep{JiangQiao_TIF_QAHE_PRB2012, Chang_TIF_ferromagnetism_AHE_AM2013, Chang_QAHE_exper_Science2013}. Hybrid structures such as TI/superconductor junctions have been fabricated \citep{Sochnikov_TISC_nano_2013,Jin-Feng_TISC_2014}, which are expected to give rise to topological superconductivity and Majorana fermions \citep{SDS_TQC_RMP08,Qiang-Hua_TISC_2014_SciRep}.

Transport experiments and theoretical work have mostly focused on longitudinal \citep{Adam_2D_Tran_prb_2012,Qiuzi_2Dtran_phimu_2012_prb,JMShao-oscillation,Costache_prl_2014,Kozlov_prl_2014,Syers_SmB_prl_2015} and Hall transport properties \citep{Chang_TIF_ferromagnetism_AHE_AM2013,Kim-Hall2013,Jing_Shoucheng_QAHE_2014_prb,ShiJunren_QAHE_prl_2014}, thermoelectric response \citep{Dohun_Thermoelectric_nl_2014,Durst_2015,ZhangJinsong_prb_2015} and weak antilocalization \citep{He_Bi2Te3_Film_WAL_ImpEff_PRL11,Cheng_TI_STM_LL_PRL10,Tkachov_HgTe_WAL_PRB11,Competition_WL_WAL_2011}, all essentially single-particle phenomena. The interplay of strong spin-orbit coupling and electron-electron interactions in TIs is at present not completely understood \citep{Culcer_TI_Int_PRB11, Yamaji-PRB-2011,Peters-PRL-2012,Yoshida-PRB-2012,Ostrovsky_TI_IntCrit_PRL10,WangCulcer_TI_Kondo_PRB2013,
LiQiuzi_2013_prb,Hai-Zhou_Conductivity_2014_prl}.

An interaction effect that can be tested experimentally in transport is Coulomb drag, which is caused by the transfer of momentum between electrons in different layers due to the interlayer electron-electron scattering. Coulomb drag has been used for decades as an experimental probe of interactions \citep{1977_drag, 1995_Hall_drag,1999_Hall_drag}, and has recently attracted considerable attention in massless Dirac fermion systems such as graphene \citep{Tse_SO_Drag_PRB07, Katsnelson2011, Hwang2011, Tse2007, Amorim_drag, M.Carrega2012, Kim2011, Narozhny_drag_2012, Kim2012, Gorbachevi_nature_2012, Sch_drag_prl_2013, Gamucci_nature_2014, Hall_drag_Graphene,Song_Halldrag_2013}. Our focus in this paper is on Coulomb drag in TIs with no magnetic impurities. Unlike graphene, the spin and orbital degrees of freedom are coupled by the strong spin-orbit interaction, TIs have an odd number of valleys on a single surface, and the relative permittivity is different, while in known band TIs screening is qualitatively and quantitatively different, since it does not involve the interplay of the layer and valley degrees of freedom. All these features impact the drag current. We introduce a density matrix method to calculate the Coulomb drag current in topological insulator films, which fully takes into account the spin degree of freedom and interband coherence. The central result of our work is the drag resistivity, which analytically takes the form
 \begin{equation}\label{rhoD}
 \rho_\text{D}=-\frac{\hbar}{e^2}\frac{\zeta(3)}{16\pi}\frac{(k_\text{B}T)^2}{A^2r^2_sn^{\frac{3}{2}}_\text{a}
 n^{\frac{3}{2}}_\text{p}d^4},
\end{equation}
where $k_\text{B}$ is the Boltzmann constant, $A$ is the TI spin-orbit constant,  $r_s$ is the Wigner-Seitz radius (effective fine structure constant) which represents the ratio of the electrons' average Coulomb potential and kinetic energies, $d$ is the layer separation and $n_{\text{a},\text{p}}$ are the electron densities in the active and passive layers, respectively. For a single-valley system $r_s = e^2/(2\pi\epsilon_0\epsilon_r A)$, with $\epsilon_r$ the relative permittivity. The intralayer resistivity $\rho_{\text{a},\text{p}}=\frac{4\pi\hbar^2}{e^2Ak_{\text{F}_{\text{a},\text{p}}}\tau_{\text{a},\text{p}}}$ with $k_{\text{F}_{\text{a},\text{p}}}$ the Fermi wave vectors.

The outline of this paper is as follows. In Sec.~\ref{sec:electron-electron} the interlayer electron-electron scattering matrix is given, including the interlayer screened Coulomb interaction. In Sec.~\ref{sec:Kinetic} we derive the kinetic equation of  topological insulators for spin density matrices of top and bottom surfaces with the full scattering term in  presence of an arbitrary elastic scattering potential to linear order in the impurity density. In Sec.~\ref{sec:drag-current},  we calculate the analytical and numerical  expressions of drag resistivity. Our findings are summarized in Sec.~\ref{sec:disc}, and we also discusses the broader implications of our results and presents a comparison with graphene. Sec.~\ref{sec:ext} discusses extensions of our theory to treat the multi-valley case and ultra-thin films, and briefly touches upon exciton condensation. Finally, Sec.~\ref{sec:con} contains our conclusions.

\section{Electron-electron interaction} \label{sec:electron-electron}

The system is described by the many-particle density matrix $\hat{F}$, which obeys the quantum Liouville equation \cite{Vasko}
\begin{equation}
\frac{\text{d}\hat{F}}{\text{d}t}+\frac{i}{\hbar}[\hat{H},\hat{F}]=0,
\end{equation}
where $\hat{H}=\hat{H}^{1e}+\hat{V}^{ee}$ with
\begin{equation}\label{H+V}
\ba
&\dps \hat{H}^{1e}\!=\!\sum_{\alpha\beta}H_{\alpha\beta}c^{\dag}_{\alpha}c_{\beta},\\[3ex]
&\dps \hat{V}^{ee}\!=\!\frac{1}{2}\sum_{\alpha\beta\gamma\delta}V^{ee}_{\alpha\beta\gamma\delta}c^\dag_\alpha c^\dag_\beta c_\gamma c_\delta.
\ea
\end{equation}
In a two-layer system the indices $\alpha\equiv{\bm k}s_{\bm k}l$ represent the wave vector, band, and layer indices respectively. The band index $s_{\bm k}=\pm$ with $+$ representing the conduction band and $-$ the valence band, while the layer index $l=~(\text{a},\text{p})$ with `a' the active layer and `p' the passive layer.
The two-particle matrix element $V^{ee}_{\alpha\beta\gamma\delta}$ in a basis spanned by a generic set of wave functions $\{\phi_{\alpha}({\bm r})\}$ is given by
\begin{equation}\label{Vee}
V^{ee}_{\alpha\beta\gamma\delta}=\int \!d{\bm r}\int \!d{\bm r}' \ \phi^*_{\alpha}({\bm r})\phi^*_{\beta}({\bm r}')V^{ee}_{{\bm r}-{\bm r}'}\phi_{\delta}({\bm r})\phi_{\gamma}({\bm r}'),
\end{equation}
where $V^{ee}_{{\bm r}-{\bm r}'}=\frac{e^2}{4\pi \epsilon_0\epsilon_r|{\bm r}-{\bm r}'|}$ is the unscreened Coulomb interaction.

The one-particle reduced density matrix is the trace
\begin{equation}
\rho_{\xi\eta}=\text{tr}(c^\dag_\eta c_\xi \hat{F})\equiv\langle c^\dag_\eta c_\xi\rangle\equiv\langle \hat{F}\rangle_{1e},
\end{equation}
 which satisfies \cite{Culcer_TI_Int_PRB11}
 \begin{equation}\label{kinetic_original_0}
 \frac{\text{d}\rho_{\xi\eta}}{\text{d}t}+\frac{i}{\hbar}[\hat{H}_{1e},\hat{\rho}]_{\xi\eta}=\frac{i}{\hbar}\langle[\hat{V}_{ee},c^\dag_\eta c_\xi]\rangle,
 \end{equation}
 where the many-electron averages such as $\langle[\hat{V}_{ee},c^\dag_\eta c_\xi]\rangle$ are factorized as
\begin{equation}\label{correlation}
\langle c^\dag_\alpha c^\dag_\beta c_\gamma c_\delta\rangle\!=\!\langle c^\dag_\alpha c_\delta\rangle\langle c^\dag_\beta c_\gamma\rangle\!-\!\langle c^\dag_\alpha c_\gamma\rangle\langle c^\dag_\beta c_\delta\rangle\!+\! G_{\alpha\beta\gamma\delta}.
\end{equation}
 in which we introduce the $G_{\alpha\beta\gamma\delta}$ as the matrix elements of the two-particle correlation operator $\hat{G}$. $G_{\alpha\beta\gamma\delta}$ give rise to the electron-electron scattering term in the kinetic equation \citep{Vasko}. The first two terms on the right side of the Eq.~(\ref{correlation}) which represent the Hartree-Fock mean-field part of the electron-electron interactions have been investigated in Ref.~\citep{Culcer_TI_Int_PRB11}. In Ref.~\citep{Culcer_TI_Int_PRB11} it was demonstrated that the electrical current and nonequilibrium spin polarization undergo a small renormalization due to the mean-field part of  electron-electron interactions and are consequently slightly reduced as compared with their non-interacting values. We are not including this weak renormalization here, so the right-hand side of Eq.~(\ref{kinetic_original_0}) only gives the electron-electron scattering term $\hat{J}_{ee}(\hat{\rho}|t)$ which has two contributions, representing intralayer and interlayer electron-electron scattering. Moreover, since the intralayer electron-electron scattering does not contribute to the drag current, we concentrate on the interlayer electron-electron scattering, for which the scattering term is denoted by $J^{\text{Inter}}(\hat{\rho}|t)$. We use below the basis of the eigenstate problem and account for only diagonal part of the density matrix $ J^{\text{Inter}}(f_{\bm k})=\langle{\bm k}|J^{\text{Inter}}(\hat{\rho}|t)|{\bm k}\rangle$
\begin{equation}\label{electron-electron-m}
\ba
&\dps J^{\text{Inter}}(f_{\bm k})\!=\!\langle{\bm k}|\frac{1}{\hbar^2L^4}\!\sum_{{\bm q}{\bm q}_1}v_qv_{q_1}\!\int^{\infty}_0\text{d}t_1\text{e}^{\lambda t_1}\Big[\text{e}^{-i{\bm q}\cdot{\bm r}},\\[3ex]
&\dps \hat{S}(t,t_1)(\mathds{1}-\hat{\rho}_{t_1})\text{e}^{i{\bm q}_1\cdot{\bm r}}\hat{\rho}_{t_1}
\hat{S}^+(t,t_1)\big\{\big\{
\text{e}^{i{\bm q}\cdot{\bm r}}\hat{S}(t,t_1)\text{e}^{-i{\bm q}_1\cdot{\bm r}}\hat{\rho}_{t_1}\\[3ex]
&\dps \hat{S}^+(t,t_1)\big\}\big\}\!-\!\hat{S}(t,t_1)\hat{\rho}_{t_1}\text{e}^{i{\bm q}_1\cdot{\bm r}}(\mathds{1}-\hat{\rho}_{t_1})\hat{S}^+(t,t_1)\big\{\big\{
\text{e}^{i{\bm q}\cdot{\bm r}}\\[3ex]
&\dps  \hat{S}(t,t_1)\hat{\rho}_{t_1}\text{e}^{-i{\bm q}_1\cdot{\bm r}}\hat{S}^+(t,t_1)\big\}\big\}+\hat{S}(t,t_1)[\hat{\rho}_{t_1},\text{e}^{i{\bm q}_1\cdot{\bm r}}]\\[3ex]
&\dps \hat{S}^+(t,t_1)\big\{\big\{\text{e}^{i{\bm q}\cdot{\bm r}}
\hat{S}(t,t_1)\hat{\rho}_{t_1}\text{e}^{-i{\bm q}_1\cdot{\bm r}}\hat{\rho}_{t_1}\hat{S}^+(t,t_1)\big\}\big\}\Big]|{\bm k}\rangle,
\ea
\end{equation}
where $v_q=\frac{e^2}{2\epsilon_0\epsilon_r q}$, $\mathds{1}$ is the identity matrix, $L^2$ the area of the 2D system, $\hat{S}(t,t_1)$ the time evolution operator and $\{\{\hat{A}\}\}\equiv\hat{A}-\text{tr}\hat{A}$ \citep {Vasko}. The momentum transfer ${\bm q}={\bm q}_1={\bm k}-{\bm k}_1={\bm k}'_1-{\bm k}'$. Following a series of simplifications, the interlayer Coulomb interaction eventually takes the form $v^{(\text{pa})}_{|{\bm k}-{\bm k}_1|}$. Without screening $v^{(\text{pa})}_{|{\bm k}-{\bm k}_1|}=v_q\text{e}^{-qd}$. To account for screening, we employ the standard procedure of solving the Dyson equation for the two-layer system in the random phase approximation (RPA) discussed in Ref.~\citep{1995_Hall_drag}. In this approach, $v^{(\text{pa})}_{|{\bm k}-{\bm k}_1|}$ in Eq.~(\ref{electron-electron-m}) becomes the dynamically screened interlayer Coulomb interaction
\begin{equation}
V({\bm q},\omega)=\frac{v_q\text{e}^{-qd}}{\epsilon({\bm q},\omega)}.
\end{equation}
The dielectric function of the coupled layer system is
\begin{equation}\label{dielectric}
\ba
\epsilon({\bm q},\omega) \!&\dps \!=[1-v_q\Pi_\text{a}({\bm q},\omega)][1-v_q\Pi_\text{p}({\bm q},\omega)]\\[3ex]
&\dps -[v_q\text{e}^{-qd
}]^2\Pi_\text{a}({\bm q},\omega)\Pi_\text{p}({\bm q},\omega),
\ea
\end{equation}
in which the polarization function is obtained by summing the lowest bubble diagram and takes the form
\begin{equation}
\Pi_{l}({\bm q},\omega)\!=\!-\frac{1}{L^2}\sum_{{\bm k}ss'}\frac{F^{(l)}_{s_{{\bm k}}s_{{\bm k}'}}(f^{(l)}_{0{\bm k},s}-f^{(l)}_{0{\bm k}',s'})}
{\varepsilon^{(l)}_{{\bm k},s}\!-\!\varepsilon^{(l)}_{{\bm k}',s'}\!+\!\hbar \omega\!+\!i0^{+}},
\end{equation}
 with $f^{(l)}_{0{\bm k},s}\equiv f^{(l)}_0(\varepsilon_{{\bm k}s})$ the equilibrium Fermi distribution function, and
 $F^{(l)}_{s_{\bm k}s_{{\bm k}_1}}=\langle s_{\bm k}l|s_{{\bm k}_1}l\rangle\langle s_{{\bm k}_1}l|s_{\bm k}l\rangle$  the wavefunction overlap. In topological insulator with no tunneling

 \begin{equation}\label{overlap}
 F^{(l)}_{s_{\bm k}s_{{\bm k}_1}}=\frac{1}{2}(1+ss'\frac{k+q\cos\phi}{|{\bm k}+{\bm q}|}).
 \end{equation}

In analytical calculations, the dynamical screened Coulomb interaction $V({\bm q},\omega)$ is usually replaced by the static screened $V_q$ at low temperatures,
 $\Pi_{l}(q,0)=-\frac{k_{\text{F}_l}}{2\pi A}$ with $k_{\text{F}_l}$ the Fermi wave vector. This is because the typical frequencies contributing to the integral are only of the order of $k_\text{B}T/\hbar$. We approximate Coulomb interaction with $k_\text{F}d\gg1$ as
\begin{equation}\label{V-static}
|V_q|^2=\bigg(\frac{ e^2}{2\epsilon_0\epsilon_r}\bigg)^2\frac{q^2}{4k^2_{\text{TF}_\text{a}}k^2_{\text{TF}_\text{p}}\sinh^2(qd)},
\end{equation}
where $k_{\text{TF}_l}=\frac{r_s k_{\text{F}_l}}{2}$  is the Thomas-Fermi wave vector and  $r_s$ is the Wigner-Seitz radius introduced in Eq.~(\ref{rhoD}).

\section{Kinetic Equations}\label{sec:Kinetic}

At this stage one may include explicitly disorder and driving electric field in the one particle Hamiltonian and write $\hat{H}^{1e}=\hat{H}_0+\hat{H}_{E}+\hat{U}$, where $\hat{H}_0$ is the band Hamiltonian of TIs, $\hat{H}_E=e\hat{\bm E}\cdot\hat{\bm r}$ is the electrostatic potential due to the driving electric flied with $\hat{\bm r}$ is a position operator and $\hat{U}$ is the disorder potential. According to Sec.~\ref{sec:electron-electron}, the quantum Liouville equation for the reduced density operator $\hat{\rho}$ satisfies
\begin{equation}
\frac{\text{d}\hat{\rho}}{\text{d}t}+\frac{i}{\hbar}[\hat{H}^{1e},\hat{\rho}]+\hat{J}_{ee}(\hat{\rho}|t)=0
\end{equation}
and will be projected onto the time-independent basis $|{\bm k},s_{\bm k},l\rangle$. Below we do not write the band indices explicitly. Since the current operator is diagonal in wave vector, the quantity of interest in determining the charge current is the part of the density matrix which is diagonal in wave vector \citep{Culcer_TI_PhysE12}, which here we denote by $f_{\bm k}$. In the absence of interlayer tunneling, the $4\times4$ matrix $f_{\bm k}$, which describes the two-layer system, is assumed to be approximately diagonal in the layer index and can be reduced to two $2\times2$ density matrices $f^{(\text{a})}_{\bm k}$ and $f^{(\text{p})}_{\bm k}$. The quantum Liouville equation can then be broken down into two separate equations, one for $f^{(\text{a})}_{\bm k}$ and one for $f^{(\text{p})}_{\bm k}$, with different driving terms. The electric field, which is only applied to the active layer, is ${\bm E}_a=E_a\hat{\bm x}$. We regard Eq.~(\ref{electron-electron-m}) as the driving term for the passive layer, since the only quantity coupling the two layers is the interlayer electron-electron scattering term. The kinetic equations take the form
\begin{subequations}\label{eqab}
 \begin{equation}\label{active}
  \frac{\text{d}f^{(\text{a})}_{\bm k}}{\text{d}t}+\frac{i}{\hbar}[H^{(\text{a})}_{0{\bm k}},f^{(\text{a})}_{\bm k}]+\hat{J}_{0}(f^{(\text{a})}_{\bm k})=-\frac{i}{\hbar}[H^E_{{\bm k}},f^{(\text{a})}_{0{\bm k}}],
 \end{equation}
\begin{equation}\label{passive}
\frac{\text{d}f^{(\text{p})}_{\bm k}}{\text{d}t}+\frac{i}{\hbar}[H^{(\text{p})}_{0{\bm k}},f^{(\text{p})}_{\bm k}]+\hat{J}_{0}(f^{(\text{p})}_{\bm k})=-J^{\text{Inter}}_{{\bm k},\text{p}},
\end{equation}
\end{subequations}
where the band Hamiltonian $H^{(l)}_{0{\bm k}}=A{\bm \sigma}\cdot({\bm k}\times\hat{\bm z})\equiv-Ak{\bm \sigma}\cdot\hat{\bm \theta}$, with $\hat{\bm \theta}$ the tangential unit vector corresponding to ${\bm k}$. The projection of $J^{\text{Inter}}(f_{\bm k})$ onto the eigenstates of the passive layer is denoted by $J^{\text{Inter}}_{{\bm k},\text{p}}$, where
\begin{equation}\label{electron-electron-mm}
 \ba
 J^{\text{Inter}}_{{\bm k},\text{p},s_{\bm k}}&\dps\!=\!-\frac{2\pi}{\hbar L^4}\!\sum_{{\bm k}_1{\bm k}'{\bm k}'_1}\!|v_{|{\bm k}-{\bm k}_1|}^{(\text{pa})}|^2\delta_{{\bm k}+{\bm k}',{\bm k}_1+{\bm k}'_1}F^{\text{(p)}}_{s_{\bm k}s_{{\bm k}_1}} \\[3ex]
&\dps \times F^{\text{(a)}}_{s_{{\bm k}'}s_{{\bm k}'_1}}\delta[\varepsilon^{(\text{p})}_{k_1,s_{{\bm k}_1}}\!\!-\!\varepsilon^{(\text{p})}_{k,s_{\bm k}}\!+\!\varepsilon^{(\text{a})}_{k'_1,s_{{\bm k}'_1}}\!-\!\varepsilon^{(\text{a})}_{k',s_{\bm k'}}\!]\\[3ex]
&\dps \times\big\{f^{(\text{p})}_{{\bm k},s_{\bm k}}[1\!-\!f^{(\text{p})}_{{\bm k}_1,s_{{\bm k}_1}}]f^{(\text{a})}_{{\bm k}',s_{\bm k'}}[1\!-\!f^{(\text{a})}_{{\bm k}'_1,s_{{\bm k}'_1}}]\\[3ex]
&\dps -[1\!-\!f^{(\text{p})}_{{\bm k},s_{\bm k}}]f^{(\text{p})}_{{\bm k}_1,s_{{\bm k}_1}}[1\!-\!f^{(\text{a})}_{{\bm k}',s_{\bm k'}}]f^{(\text{a})}_{{\bm k}'_1,s_{{\bm k}'_1}}\!\big\}.
\ea
\end{equation}
We do not need to consider the interlayer electron-electron collision integral $J^{\text{Inter}}_{{\bm k},\text{a}}$ in the active layer, since that does not produce a drag current. The electron-impurity scattering term in the first Born approximation is given by \citep{Culcer_TI_AHE_PRB11}
\begin{equation}\label{J0}
\ba
\hat{J}_{0}(f^{(l)}_{\bm k})\!=\!\big\langle\int^\infty_0\frac{\text{d}t'}{\hbar^2}[\hat{U},e^{-i\hat{H}t'/\hbar}[\hat{U},\hat{f}]
e^{i\hat{H}t'/\hbar}]\big\rangle_{{\bm k}{\bm k}},
\ea
\end{equation}
where the notation $\langle\rangle$ denotes the average over impurity configurations.

We will write $f^{(l)}_{\bm k}=n^{(l)}_{\bm k}\mathbb{1}+S^{(l)}_{\bm k}$, with $S^{(l)}_{\bm k}$ a $2\times2$ Hermitian matrix which can be written in terms of the Pauli spin matrices. Every matrix in this section can be written in terms of a scalar part, labeled by the subscript $n$, and two spin-dependent parts $\sigma_{{\bm k}\parallel}$ and $\sigma_{{\bm k}\perp}$. The kinetic Eq.~(\ref{eqab}) can be written as
\begin{subequations}\label{decomposation}
\begin{equation}
 \frac{\text{d} n^{(l)}_{\bm k}}{\text{d}t}+P_n\hat{J}(f^{(l)}_{\bm k})=\mathcal{D}^{(l)}_{{\bm k},n},
\end{equation}
\begin{equation}\label{parallel}
\frac{\text{d}S^{(l)}_{{\bm k},\parallel}}{\text{d}t}+P_{\parallel}\hat{J}(f^{(l)}_{\bm k})=\mathcal{D}^{(l)}_{{\bm k},\parallel},
\end{equation}
\begin{equation}\label{perp}
\frac{\text{d}S^{(l)}_{{\bm k},\perp}}{\text{d}t}+\frac{i}{\hbar}[H^{(l)}_{0\bm k},S^{(l)}_{{\bm k},{\perp}}]+P_{\perp}\hat{J}(f^{(l)}_{\bm k})
=\mathcal{D}^{(l)}_{{\bm k},\perp},
\end{equation}
\end{subequations}
where $\mathcal{D}^{(l)}_{\bm k}$ is the driving term for layer $l$. For the active layer, $\mathcal{D}^{(\text{a})}_{\bm k}=-\frac{i}{\hbar}[H^{E}_{\bm k}, f^{(\text{a})}_{0{\bm k}}]$; for the passive layer, $\mathcal{D}^{(\text{p})}_{\bm k}=J^{\text{Inter}}_{{\bm k},\text{p}}$, which both have three components $\mathcal{D}^{(l)}_{{\bm k},n}$, $\mathcal{D}^{(l)}_{{\bm k},\parallel}$ and $\mathcal{D}^{(l)}_{{\bm k},\perp}$. The projection operator $P_{\parallel}$ acts on a matrix $\mathcal{M}$ as $\text{tr}(\mathcal{M}\sigma_{{\bm k}\parallel})$, where tr refers to the matrix (spin) trace. Analogous definitions hold for the operators $P_{\perp}$ and $P_n$. The operators $P_\parallel$, $P_{\perp}$, and $P_n$ single out the parts of the density matrix which are parallel to $H^{(l)}_{0{\bm k}}$ (in matrix language), orthogonal to $H^{(l)}_{0{\bm k}}$, and scalar, respectively.

\section{Calculation of Coulomb drag }\label{sec:drag-current}

To obtain the  drag resistivity, the kinetic equation for the passive layer must be solved.
We feed the equilibrium density matrix for the passive layer $f^{(\text{p})}_{\bm k}=n^{(\text{p})}_{{\bm k},0}+S^{(\text{p})}_{{\bm k},0}$ and the full density matrix of the active layer $f^{(\text{a})}_{\bm k}=n^{(\text{a})}_{{\bm k},0}+S^{(\text{a})}_{{\bm k},0}+\delta f^{(\text{a})}_{E{\bm k}}$ into Eq.~(\ref{electron-electron-m}), where $ \delta f^{(\text{a})}_{E{\bm k}} = n^{(\text{a})}_{E{\bm k}}+S^{(\text{a})}_{E{\bm k}}$ is  a small correction to the distribution function caused by the applied electric field in the active layer
\begin{subequations}
\begin{equation}
n^{(\text{a})}_{E{\bm k}}=\frac{e{\bm E}_\text{a}\tau_\text{a}(k)\cdot\hat{{\bm k}}}{2\hbar}\pd{(f^{(\text{a})}_{0+}+f^{(\text{a})}_{0-})}{k},
\end{equation}
\begin{equation}
 S^{(\text{a})}_{E{\bm k},\parallel}=\frac{e{\bm E}_\text{a}\tau_\text{a}(k)\cdot\hat{{\bm k}}}
{2\hbar}\pd{(f^{(\text{a})}_{0+}-f^{(\text{a})}_{0-})}{k}\sigma_{{\bm k}\parallel},
\end{equation}
\end{subequations}
with $\tau_\text{a}(k)$ the momentum scattering time \citep{Culcer_TI_PhysE12}.
\om \begin{equation}
\ba
&\dps \frac{1}{\tau_\text{a}(k)}=\frac{\varepsilon_+}{A}\frac{n_iW_k}{2A\hbar}(\varsigma_0-\varsigma_1),\quad
 W_k=\frac{Z^2e^4}{4\epsilon^2_0\epsilon^2_rk^2_\text{F}},
 \ea
\end{equation}
 and $\varsigma_{0,1}$ refer to the Fourier components of
 \begin{equation}
 \varsigma=\frac{1}{2}(1+\cos\gamma)\frac{1}{(\sin\frac{\gamma}{2}+\frac{k_{\text{TF}}}{2k_\text{F}})^2}.
\end{equation}\fi
 The resulting interlayer electron-electron collision integral becomes the driving term for the passive layer, and we search for the solution of the kinetic equation for the passive layer, which will yield the drag current. It is easy to verify that the electron-electron collision integral vanishes if we replace the density matrix of both layers by its equilibrium form $f^{(l)}_{0{\bm k}} = n^{(l)}_{0{\bm k}}+S^{(l)}_{0{\bm k}}$. Here $n^{(l)}_{0{\bm k}}=\big[f^{(l)}_{0{\bm k},+}+f^{(l)}_{0{\bm k},-}\big]/2$ and $S^{(l)}_{0{\bm k}}=\big[f^{(l)}_{0{\bm k},+}-f^{(l)}_{0{\bm k},-}\big]\sigma_z/2$ are the equilibrium distribution for the charge and spin dynamics in layer $l$ respectively, where  $f^{(l)}_{0{\bm k},\pm}$ are the Fermi-Dirac functions for the two energy eigenstates. We solve the kinetic Eq.~(\ref{parallel}), yielding the density matrix for the passive layer in the steady state
\begin{equation}
\ba
 S^{(\text{p})}_{{\bm k},\parallel}\!&\dps=\frac{e\pi\sigma_{{\bm k}\parallel}}{4k_\text{B}TL^4}\sum_{{\bm k}'{\bm q}}\int \text{d} \omega\frac{|V({\bm q},\omega)|^2}{\sinh^2\frac{\beta\hbar\omega}{2}}\\[3ex]
&\dps \times F^{++}_{{\bm k},{\bm k}_1}\delta[\varepsilon^{(\text{p})}_{k_1,+}-\varepsilon^{(\text{p})}_{k,+}+\hbar \omega](f^{(\text{p})}_{0{\bm k},+}-f^{(\text{p})}_{0{\bm k}_1,+})\\[3ex]
&\dps \times F^{++}_{{\bm k},{\bm k}'_1} \delta[\varepsilon^{(\text{a})}_{k'_1,+}-\varepsilon^{(\text{a})}_{k',+}-\hbar\omega](f^{(\text{a})}_{0{\bm k}',+}-f^{(\text{a})}_{0{\bm k}'_1,+})\\[3ex]
 &\dps \times{\bm E}_{\text{a}}\cdot\big[\tau_\text{a}(k'_1){\bm v}_{{\bm k}'_1}-\tau_\text{a}(k'){\bm v}_{{\bm k}'}\big]\tau_\text{p}(k),
\ea
\end{equation}
where ${\bm v}_{\bm k}=\frac{Aa_k}{\hbar}\hat{\bm k}$.
The current operator due to the band Hamiltonian is $
\hat{\bm j}=\frac{eA}{\hbar}{\bm \sigma}\times\hat{\bm z}$. With ${\bm j}_\text{p}=\text{tr}(\hat{\bm j}S^{(\text{p})}_{{\bm k},\parallel})={\bm j}_\text{p}=\sigma_\text{D}{\bm E}_{\text{a}}$, we have
\begin{equation}\label{cll}
\ba
\sigma_\text{D}\!=\!\frac{e^2}{16\pi k_\text{B}T}\!\sum_{\bm q}\!\int \text{d}\omega\frac{|V({\bm q},\omega)|^2\text {Im}[\chi^{++}_\text{a}({\bm q},
\omega)]\text {Im}[\chi^{++}_\text{p}({\bm q},
\omega)]}{\sinh^2 \frac{\beta\hbar\omega}{2}},
\ea
\end{equation}
where the nonlinear drag susceptibility for the conduction band of one layer is given by
\begin{equation}
\ba
 \chi^{++}_l({\bm q},\omega)&\dps =-\frac{2\pi}{L^2}\! \sum_{\bm k}F^{++}_{{\bm k},{\bm k}+{\bm q}}(f^{(l)}_{{\bm k},+}-f^{(l)}_{|{\bm k}+{\bm q}|,+})\\[3ex]
&\dps \times\frac{[\tau_l(k){\bm v}_{\bm k}-\tau_l(|{\bm k}+{\bm q}|){\bm v}_{{\bm k}+{\bm q}}]}{\varepsilon^{(l)}_{{\bm k},+}-\varepsilon^{(l)}_{|{\bm k}+{\bm q}|,+}+\hbar\omega+i0^+}.
\ea
\end{equation}
The drag problem reduces to the calculation of the nonlinear susceptibility of the system. This fact reflects the physical mechanism behind the drag phenomenon: the drag current is a result of the rectification by the passive layer of the fluctuating electric field created by the active layer \citep{1995_Hall_drag}. In special cases when the intralayer electron-electron correlations are absent, the nonlinear susceptibility is reduced to the product of the diffusion constant and the imaginary part of the polarization operator \citep{1995_Hall_drag}. At low temperatures the predominant contribution to drag is due to intraband processes near the Fermi surface. When the Fermi level is finite, i.e. above the Dirac point, electrons take more energy to transition from the valence band to the conduction band than to transition within the conduction band or valence band, and with a small excitation energy the channel involving interband transitions becomes inaccessible \citep{Tse2007}. The small interlayer momentum transfer and excitation energy, i.e. $q<2k_\text{F}$, $\hbar\omega<Aq$  is the dominant region of polarization contributing to the drag problem. Writing $\tau_l(k)=\tau_{0}k_l$, we have for the imaginary part of the susceptibility
\begin{equation}\label{S}
\ba
&\dps \text{Im}\chi^{++}_l({\bm q},\omega)=-\frac{\tau_0q^2\hat{\bm q}}{8\pi \hbar\sqrt{1-(\frac{\hbar\omega}{Aq})^2}}\\[3ex]
&\dps \times\big[G_>\Big(\frac{2k_{\text{F}_l}-\frac{\hbar\omega}{A}}{q}\Big)- G_>\Big(\frac{2k_{\text{F}_l}+\frac{\hbar\omega}{A}}{q}\Big)\big],
\ea
\end{equation}
with
\begin{equation}\label{G}
 G_>(x)=x\sqrt{x^2-1}-\text{arccosh}(x).
\end{equation}
The susceptibility, which in general is dependent on the electron momentum, reduces to a momentum-independent form at low temperatures, with the scattering time evaluated at the Fermi level $\tau_{\text{F}_l}=\tau_0k_{\text{F}_l}$.

\section{Discussion}\label{sec:disc}

To obtain analytical results, we feed the susceptibility Eq.(\ref{S}) and the approximate Coulomb interaction Eq.(\ref{V-static}) into Eq.(\ref{cll}). With
\begin{equation}
\int^{\infty}_0\frac{\omega^2\text {d}\omega}{\sinh^2(\hbar\omega/2k_\text{B}T)}=\frac{4\pi^2}{3}\left(\frac{k_\text{B}T}{\hbar}\right)^3,
\end{equation}
we have the drag resistivity
\begin{equation}\label{ll-result}
\rho_\text{D} \approx-\frac{\sigma_{\text{D}}}{\sigma_{\text{a}}\sigma_{\text{p}}} 
=-\frac{\hbar}{e^2}\frac{\zeta(3)}{16\pi}\frac{(k_\text{B}T)^2}{A^2r^2_sn^{\frac{3}{2}}_\text{a}n^{\frac{3}{2}}_\text{p}d^4}.
\end{equation}

\begin{figure}[Top]
\begin{center}
\includegraphics[width=0.9\columnwidth]{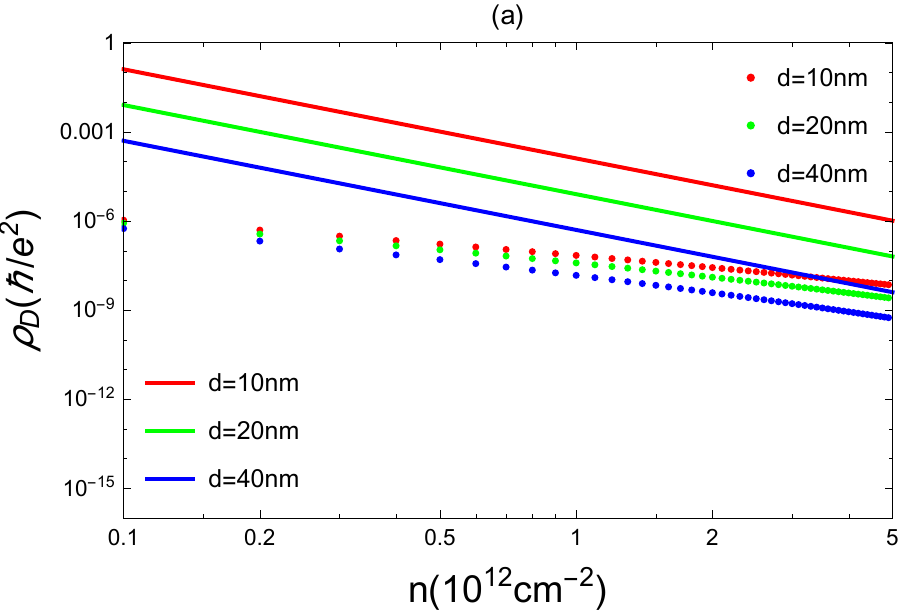}\\
\includegraphics[width=0.9\columnwidth]{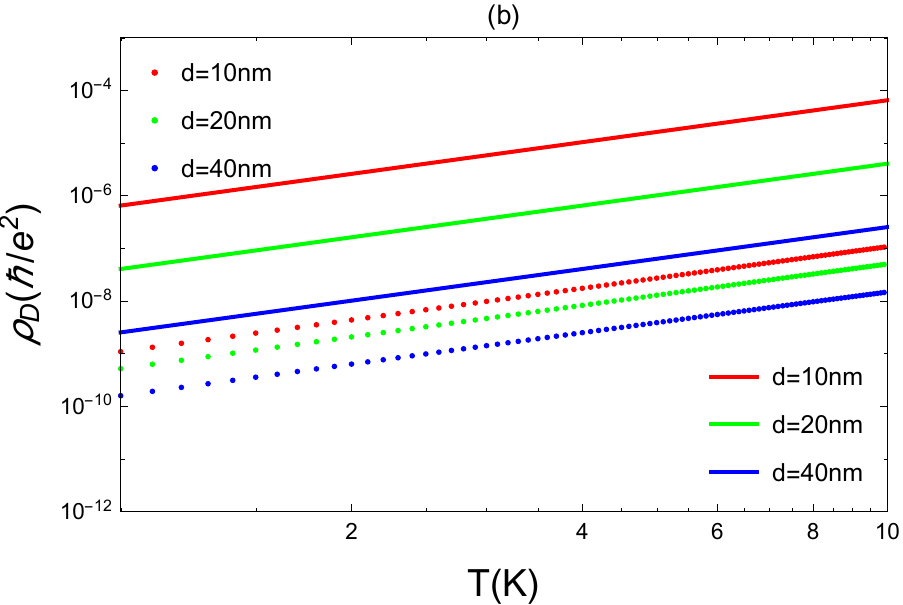}\\
\includegraphics[width=0.9\columnwidth]{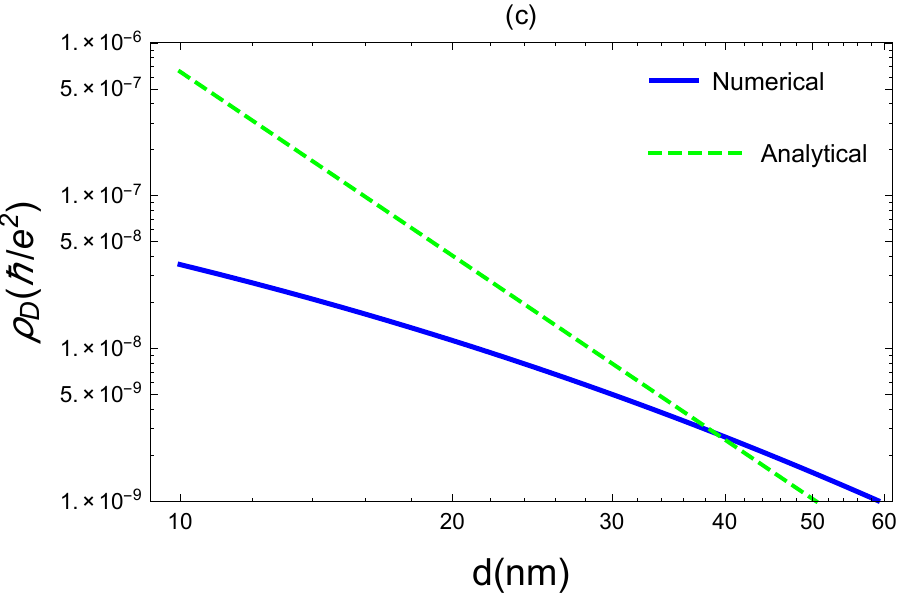}
\caption{\label{xxD} Behavior of drag resistivity $\rho_{\text{D}}$ as a function of electron concentration $n$, temperature $T$, layer separation $d$: (a) is electron density dependence of $\rho_\mathrm{D}$ at $T=5~\mathrm{K}$; (b) is temperature dependence of $\rho_\mathrm{D}$ with $k_\mathrm{F_{\mathrm{a}}}=k_\mathrm{F_{\mathrm{p}}}=0.5~\mathrm{nm}$; (c) is layer separation $d$ dependence of $\rho_\mathrm{D}$ at $T=5\text{K}$ and $k_\mathrm{F_{\mathrm{a}}}=k_\mathrm{F_{\mathrm{p}}}=0.5~\mathrm{nm}$.   Dielectric constant $\epsilon_r=50$.  Real and dotted (dashed) lines represent the numerical  and analytical  respectively.}
\end{center}
\end{figure}

In Fig.~\ref{xxD}(a) we present numerical results for the dependence of the Coulomb drag resistivity on the electron number density. The drag resistivity displays a $\frac{1}{n^{\alpha}_{\text{a}}n^{\alpha}_{\text{p}}}$ dependence with $\alpha<1.5$ for $d=10,20,40~\text{nm}$. With increasing electron density $n$ the coefficient $\alpha$ approaches $1.5$. The fact that the exact numerical results shown in Fig.~\ref{xxD}(a) disagree more strongly with the analytical results for smaller values of $n$ and $d$ is understandable, since  Eq. (\ref{ll-result}) applies only in the $k_\text{F} d\gg1$ limit with $k_\text{F}=\sqrt{4\pi n}$. This trend of an increasing quantitative failure of the asymptotic analytical drag formula for small $k_\text{F}d$ has also been noted in graphene \citep{Tse2007,Hwang2011,M.Carrega2012} and 2DEG systems \citep{1995_Hall_drag,1999_Hall_drag}. The analytical result becomes more accurate with increasing $k_\text{F}d$.

In Fig.~\ref{xxD}(b) we show the Coulomb drag resistivity as a function of temperature $T$ for three different thicknesses $d=10,20,40~\text{nm}$. The overall temperature dependence of the drag resistivity increases nearly quadratically and there is no logarithmic correction due to the absence of backscattering in TIs. The $T^2$ dependence stems from the allowed phase space where electron-electron scattering occurs at low temperature, and is expected for any interaction strength between the top and bottom layers of TIs as Fig.~\ref{xxD}(b) shows, provided that the carriers can be described using a Fermi liquid picture. In addition, in TIs the acoustic phonon velocity is smaller than in graphene. These facts make the contribution of electron-phonon scattering processes to the resistivity much more important in the surface of 3DTIs than in graphene. For the surfaces of 3DTIs the effect of electron-phonon scattering events becomes important already for $T$ as low as $10\text{K}$. For this reason we take consider temperatures up to $10\text{K}$ in our numerical calculations. In graphene this effect becomes relevant only beyond $T\gtrsim200\text{K}$. It is also evident that Eq. (\ref{ll-result}) becomes increasingly accurate and approaches the numerical results as the layer separation $d$ increases.

The behavior of the drag resistivity $\rho_{\text{D}}$ as a function of layer separation $d$ is shown in Fig.~\ref{xxD}(c) with $T=5\text{K}$ and $k_\mathrm{F_{\mathrm{a}}}=k_\mathrm{F_{\mathrm{p}}}=0.5~\mathrm{nm}$. The trend of the exact numerical results changes more slowly, a fact that is also embodied in Fig.~\ref{xxD}(a). Interestingly, in drag experiments on graphene \citep{Gorbachevi_nature_2012}, the $d$ dependence of the drag resistivity is much slower than the $1/d^4$ expected in the weakly interacting regime, varying approximately as $1/d^2$ for $d>4\text{nm}$, which is comparable to TIs.

\begin{figure}[Top]
\begin{center}
\includegraphics[width=0.9\columnwidth]{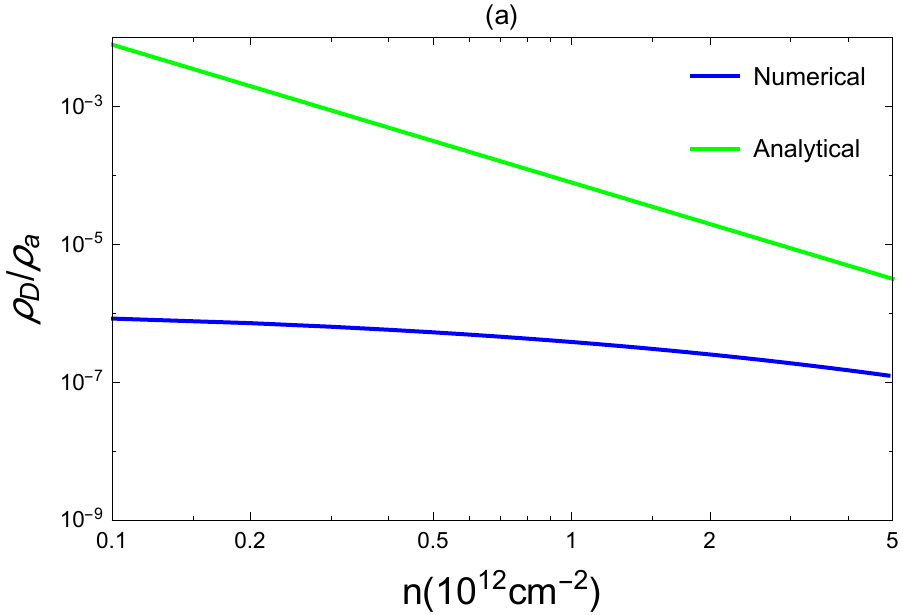}\\
\includegraphics[width=0.9\columnwidth]{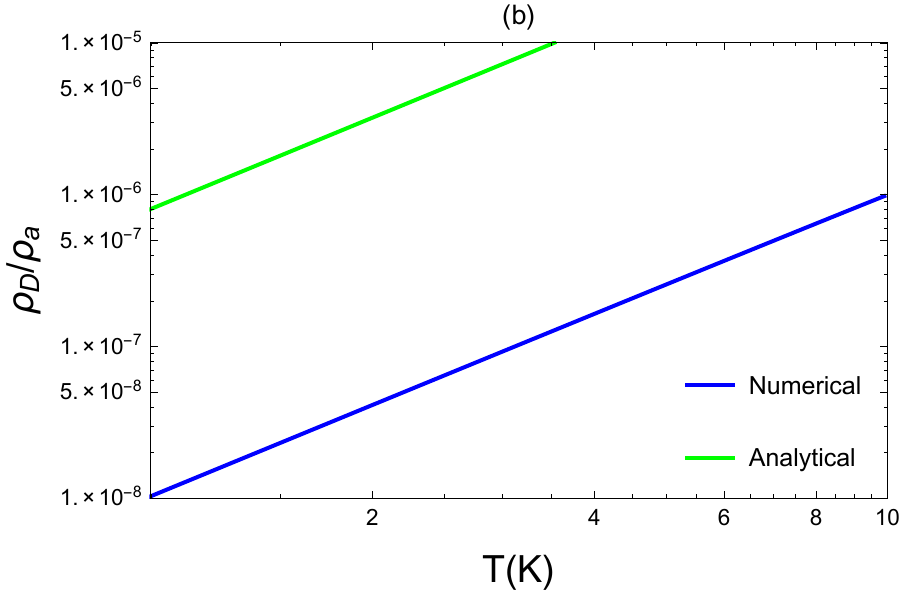}\\
\includegraphics[width=0.9\columnwidth]{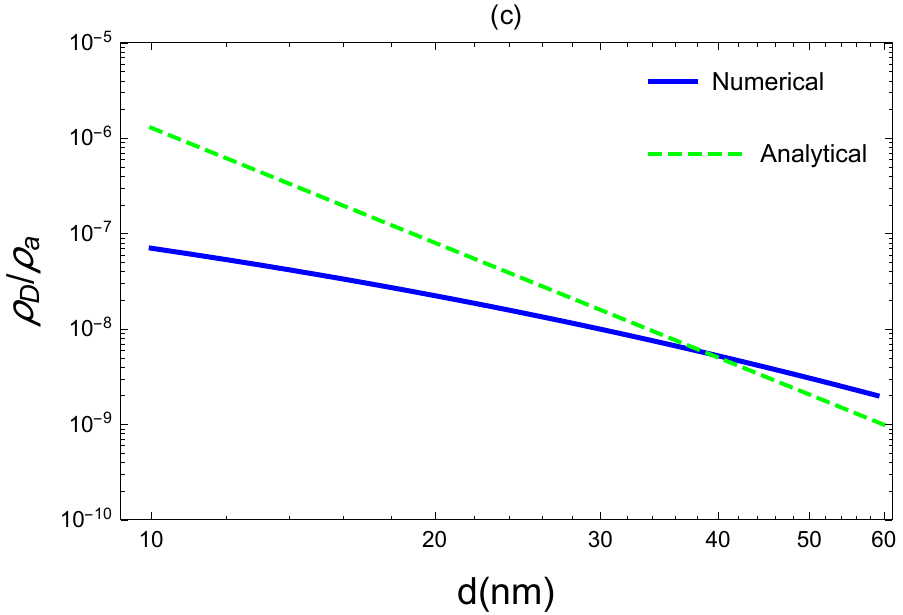}
\caption{\label{xxDxx} Numerical (blue lines) and analytical (green lines) results of ratio between drag resistivity $\rho_\mathrm{D}$ and intralayer resistivity $\rho_\text{a}$ with $\rho_\text{a}=\rho_\text{p}$:
 (a) is electron density dependence of $\frac{\rho_\mathrm{D}}{\rho_\mathrm{a}}$ at $T=5~\mathrm{K}$ and $d=20\text{nm}$; (b) is temperature dependence of $\frac{\rho_\mathrm{D}}{\rho_\mathrm{a}}$ with $k_\mathrm{F_{\mathrm{a}}}=k_\mathrm{F_{\mathrm{p}}}=0.5~\mathrm{nm}$ and $d=20\text{nm}$; (c) is layer separation $d$ dependence of $\frac{\rho_\mathrm{D}}{\rho_\mathrm{a}}$ at $T=5\text{K}$ and $k_\mathrm{F_{\mathrm{a}}}=k_\mathrm{F_{\mathrm{p}}}=0.5~\mathrm{nm}$. Dielectric constant $\epsilon_r=50$ and momentum scattering time $\tau_{\text{a,p}}=0.8~\mathrm{ps}$.}
\end{center}
\end{figure}

Fig.~\ref{xxDxx} shows the ratio between $\rho_{\text{D}}$ and the intralayer resistivity $\rho_\text{a}$ corresponding to Fig.~\ref{xxD}, which illustrates the relative magnitudes of the drag and longitudinal resistivities. The ratio is about $10^{-6}\thicksim10^{-8}$, indicating a small drag resistivity and reflecting the weak electron-electron interactions in TIs. In current 3DTIs, the dielectric constant is as large as 100, indicating weak electron-impurity and electron-electron Coulomb scattering \citep{MWWu_Peng}. So the drag resistivity of TIs is much smaller than that of graphene which the relative static dielectric constant is about 4. This places TIs squarely in the weakly-interacting regime, in which RPA-based theories are applicable.

For both Fig.~\ref{xxD} and Fig.~\ref{xxDxx}, we have only applied the approximation $k_{\text{F}}d\gg1$ for TI film thicknesses up to $6\text{nm}$, with electron densities $n\thicksim10^{12}\text{cm}^{-2}$. This is a somewhat different regime than graphene, in which the condition $k_{\text{F}}d \ll1$ can usually be satisfied. However, for other systems with small enough $k_{\text F}d$ and larger $r_s$, the interlayer Coulomb interaction is in general not small, and it remains to be determined whether it is necessary to take into account higher-order contributions in the Coulomb interaction. We have found that approximations applied to $|V_q|^2$ can lead to discrepancies between analytical and numerical results. The numerical result is a factor of $\thicksim 10^2$ lower than the analytic result, an observation which, in the case of graphene, has been confirmed by drag experiments \citep{Kim2011}.

\section{Extensions of the theory}\label{sec:ext}

The focus of the paper up to now has been on thin films of band TIs in which both layers are doped with the same type of carrier, that is, either electron-electron or hole-hole layers. Our theory can straightforwardly be extended to treat structures beyond those considered thus far. In this section we present the necessary modifications for treating the cases of multi-valley TIs and ultra-thin films, and discuss briefly the possibility of exciton condensation.

\subsection{Multi-valley case}

Certain materials, such as the topological Kondo insulator SmB$_6$, have more than one valley. In this case, a valley degeneracy factor $g_v$ will need to be introduced.
In the following, we will fix the electron density $n$. The polarisation $\Pi \propto \sqrt{g_{v}},$ while the susceptibility $\chi \propto g_{v}$, thus the screening function
$\epsilon(q,\omega)$ is proportional to $g_{v}$ \cite{SDS_Gfn_RMP11}. The drag current, and hence the drag conductivity, will remain the same as in the single-valley case. For the drag resistivity, based on Eq. (\ref{ll-result}) we will get the linear $g_{v}$ dependence because $\sigma_{a,p} \propto 1/\sqrt{g_{v}}.$ Note that there will be an intervalley
impurity scattering contribution if short-range disorder is present in the system, but this will simply result in a renormalisation of the scattering time tau by an intervalley scattering term.

\subsection{Ultra-thin films}

When the TI film is ultra thin tunnelling is enabled between the top and bottom surfaces. The tunnelling between the surface states on the top and bottom surfaces may open an energy gap in the energy spectrum \cite{Weizhe_TITF_2014_prb}. In this case, the massless Dirac Hamiltonian $H_{0{\bm k}}^{(l)}$ needs to be augmented by a series of tunnelling terms, and is generally written as:
\begin{equation}
    H_{\bm k}=A\tau_z\otimes[{\bm\sigma}\cdot({\bm k}\times\hat{\bm z})] + t\tau_x\otimes \mathbb{1},
\end{equation}
where $\tau$ matrices represent the layer pseudospin space, with $\tau_z=1$ symbolising the up surface and $\tau_z=-1$ the bottom surface. Here $\hat{\bm z}$ is the unit vector in the direction of $\bm z$, and the term $t$ represents the tunnelling matrix element between two opposite topological surfaces. After the direct diagonalization, the energy spectrum of the TI thin film is given by $\epsilon_{\bm k}=\pm\sqrt{t^2+A^2k^2}$, which has a gap of size $2t$. We discuss how this gap affects the Coulomb drag between the two surfaces.

We begin with some general results. The scattering term is:
\begin{equation}\label{Jt}
    \ba
    \dps\hat{J}(f_{\bm k})=&\dps\frac{n_i\epsilon_k}{4\pi A^2\hbar}\int_0^{2\pi} \, d\theta_{\bm k'} \, |U_{{\bm k}{\bm k'}}|^2(f_{\bm k} - f_{\bm k'}) \\[3ex]
    &\dps\times\bigg(1 + \frac{t^2}{t^2+A^2k^2} + \frac{A^2k^2\cos\gamma}{t^2+A^2k^2}\bigg),
    \ea
\end{equation}
where $\gamma = \theta_{\bm k} - \theta_{\bm k'}$ is the angle between the incident and scattered wave vectors. Note that the density matrix entering this term is the full density matrix $f_{\bm k}$ of the double-layer system, rather than its projection onto each individual layer. For $t=0$ Eq. (\ref{Jt}) reduces to the scattering term introduced in Eq.(\ref{J0}), 
and we recover the well-known factor of $1 - \cos\gamma$, which ensures there is no backscattering in TIs. Based on the ratio of the interlayer tunnelling $t$ and Fermi energy $\varepsilon_\text{F}$, two limiting cases can be identified: weak tunnelling $t \ll \varepsilon_\text{F}$ and strong tunnelling $t \gg \varepsilon_\text{F}$.
 So the wavefunction overlap Eq. (\ref{overlap}) with tunnelling becomes
\begin{equation}
F^{(l)}_{s_{\bm k}s'_{{\bm k}'}}=\frac{1}{2}(1+ss'\frac{t^2+A^2k(k+q\cos\phi)}{\sqrt{(t^2+A^2k^2)(t^2+A^2k'^2)}}).
\end{equation}
 We recalculate the polarization and susceptibility, then the related $G_>(x)$ in Eq. (\ref{G}) is rewritten as
\begin{equation}
    G_>(x)=x\sqrt{x^2-x^2_0}-(2-x^2_0)\text{arccosh}(\frac{x}{x_0}),
\end{equation}
where $ x_0=\sqrt{1+\frac{4t^2}{A^2q^2-\hbar^2\omega^2}}$.

For weak tunnelling, $t \ll Ak_F$, the carrier wave functions are overwhelmingly located in one of the two layers (surfaces), and the notion of Coulomb drag can be retained to a good approximation. The effect of tunnelling can be taken into account perturbatively. For example, the momentum relaxation time becomes
\begin{equation}
\frac{1}{\tau} \rightarrow \frac{1}{\tau} \, \bigg( 1 + \frac{t^2}{A^2k_\text{F}^2} \bigg).
\end{equation}
This implies that, as expected, interlayer tunnelling, which brings with it interlayer scattering, slightly decreases the momentum relaxation time. Screening is qualitatively different \cite{Weizhe_TITF_2014_prb}, but for weak $t$ it is a good approximation to retain the screening function defined in Eq.\ (\ref{dielectric}). With the same method of calculating  the $t=0$ case, we have the analytical result of the drag resistivity with tunnelling becomes
\begin{equation}
\rho_{\text{D}}(t)=\frac{\rho_{\text{D}}}{[1+(t/Ak_F)^2]^2} \approx \rho_{\text{D}} \bigg[ 1 - 2 \, \bigg( \frac{t}{A k_F} \bigg)^2 \bigg].
\end{equation}
We also recalculate the numerical result. Setting $t\approx 0.1Ak_\text{F}$ which corresponds to an interlayer separation $d\approx6~\text{nm}$, the result also shows that $\rho_{\text{D}}(t_0)$ is marginally smaller than the value that would be obtained by neglecting tunnelling.
 
When there the interlayer tunnelling is strong, such that $t \sim A k_F$, the notion of Coulomb drag is not applicable to the thin film system. In this case, the carrier wave functions are spread over the two layers, hence the picture of the Coulomb interaction causing charges in one layer to \textit{drag} charges in the other is no longer valid. The main effect of electron-electron interactions will be through the Coulomb renormalization of the conductivity \cite{Weizhe_TITF_2014_prb}. We note, however, that the film needs to be extremely thin for the tunnelling gap to be noticeable \cite{LuShan_TITF_MassiveDirac_spinphys_PRB2010}, therefore we expect realistic samples to lie in the weak tunnelling limit.

\subsection{On the possibility of exciton condensation}

When the active layer is doped with electrons and the passive layer is doped with holes, or vice-versa, exciton condensation may occur. This effect is driven by an exchange term in the interlayer Coulomb interaction. In principle, for interlayer exchange to be nonzero tunnelling also has to be nonzero, but one can think of a situation in which the tunnelling is negligible but the exchange is not. The interesting problem concerning the way exciton condensation impacts $\rho_{xx}$ has been considered in great detail in Ref.~\cite{Polini_Exciton_Drag_PRL2012}, where it was shown that the drag resistivity exhibits an upturn at low temperatures described by a logarithmic dependence on the temperature.

The theory of Ref.~\cite{Polini_Exciton_Drag_PRL2012} has recently been shown to be a good description of experimental observations in graphene \cite{Pellegrini_Exciton_NCom2014}, where an exciton condensate phase has been identified, with the critical temperature estimated at 10 - 100mK. This estimate was for a sample grown on a GaAs substrate, in which the effective dielectric constant is expected to be $\epsilon_r \approx 6$, whereas in current TI films the lowest experimentally reported $\epsilon_r \approx 30$ \cite{Beidenkopf_NP11,Kim_TI_e-ph_PRL12}, largely due to screening by the unavoidable bulk of the film. These observations suggest that the critical temperature in TI films could be at least an order of magnitude smaller than in graphene, placing it in the range 1 - 10 mK, which would make exciton condensation rather difficult to detect experimentally in currently available samples.

\section{Conclusion}\label{sec:con}

We have carried out a detailed analysis of the intra-layer and inter-layer electron-electron interactions in TIs in order to determine the Coulomb drag resistivity $\rho_\text{D}$ and devise a complete picture of Coulomb drag in these materials. We have found that $\rho_\text{D}$ is proportional to $T^2d^{-4}n^{-3/2}_{\text{a}}n^{-3/2}_{\text{p}}$ at low temperature and electron density. We have compared our results for $\rho_{\text{D}}$ with graphene, concluding that the drag effect is expected to be weaker in TIs, and that different regimes are accessible experimentally in TIs and graphene. The validity of certain analytical approximations for calculating $\rho_\text{D}$ has also been elucidated. The kinetic equation method presented in this work will be extended in a future publication to describe magnetically doped TIs, in which the anomalous Hall effect makes an important contribution to Coulomb drag.

\section*{Acknowledgments}
We are grateful to Zhenyu Zhang, Wenguang Zhu, Zhenhua Qiao, Changgan Zeng, Shun-Qing Shen, W.~K. Tse for enlightening discussions. This work was supported by the International Center for Quantum Design of Functional Materials HeFei National Laboratory for Physical Sciences at Microscale University of Science and Technology of China.
\section*{References}
\bibliography{refs_TI}

\begin{thebibliography}{10}
\expandafter\ifx\csname url\endcsname\relax
  \def\url#1{\texttt{#1}}\fi
\expandafter\ifx\csname urlprefix\endcsname\relax\def\urlprefix{URL }\fi
\expandafter\ifx\csname href\endcsname\relax
  \def\href#1#2{#2} \def\path#1{#1}\fi

\bibitem{KaneMele_QSHE_PRL05}
C.~L. Kane, E.~J. Mele, Quantum spin hall effect in graphene, Phys.\ Rev.\
  Lett. 95 (2005) 226801.

\bibitem{Hasan_TI_RMP10}
M.~Z. Hasan, C.~L. Kane, Topological insulators, Rev.\ Mod.\ Phys. 82 (2010)
  3045.

\bibitem{Qi_TI_RMP_10}
X.-L. Qi, S.-C. Zhang, Topological insulators and superconductors, Rev.\ Mod.\
  Phys. 83 (2011) 1057.

\bibitem{Moore_TRI_TI_Invariants_PRB07}
J.~E. Moore, L.~Balents, Topological invariants of time-reversal-invariant band
  structures, Phys.\ Rev.\ B 75 (2007) 121306.

\bibitem{Ando-TI}
Y.~Ando, Topological insulator materials, J. Phys. Soc. Jpn. 82 (2013) 102001.

\bibitem{Culcer_TI_PhysE12}
D.~Culcer, Transport in three-dimensional topological insulators: theory and
  experiment, Physica E 44 (2012) 860.

\bibitem{Tkachov_TI_Review_PSS13}
G.~Tkachov, E.~M. Hankiewicz, Phys.~Status Solidi~B 250 (2013) 215.

\bibitem{Yong-qing_FOP_2012}
\href{run:../../paper-list/Yong-qing_FOP_2012.pdf}{Li, Y. Q. }, K.~H. Wu, J.~R.
  Shi, X.~C. Xie, \href{http://dx.doi.org/10.1007/s11467-011-0190-3}{Electron
  transport properties of three-dimensional topological insulators}, Frontiers
  of Physics 7~(2) (2012) 165.
\newblock \href {http://dx.doi.org/10.1007/s11467-011-0190-3}
  {\path{doi:10.1007/s11467-011-0190-3}}.
\newline\urlprefix\url{http://dx.doi.org/10.1007/s11467-011-0190-3}

\bibitem{Moore-nature}
J.~E. Moore, The birth of topological insulators, Nature 464 (2010) 194.

\bibitem{Urazhdin_PRB04}
S.~Urazhdin, D.~Bilc, S.~D. Mahanti, S.~H. Tessmer, T.~Kyratsi, M.~G.
  Kanatzidis, Surface effects in layered semiconductors bi2se3 and bi2te3,
  Phys.\ Rev.\ B 69 (2004) 085313.

\bibitem{Hsieh_BiSb_QSHI_Nature08}
D.~Hsieh, D.~Qian, L.~Wray, Y.~Xia, Y.~S. Hor, R.~J. Cava, M.~Z. Hasan, A
  topological dirac insulator in a quantum spin hall phase, Nature 452 (2008)
  970.

\bibitem{Hsieh_BiSb_QmSpinTxtr_Science09}
D.~Hsieh, Y.~Xia, L.~Wray, D.~Qian, A.~Pal, J.~H. Dil, J.~Osterwalder,
  F.~Meier, G.~Bihlmayer, C.~L. Kane, Y.~S. Hor, R.~J. Cava, M.~Z. Hasan,
  Observation of unconventional quantum spin textures in topological
  insulators, Science 323 (2009) 919.

\bibitem{Xia_Bi2Se3_LargeGap_NP09}
Y.~Xia, D.~Qian, D.~Hsieh, L.~Wray, A.~Pal, H.~Lin, A.~Bansil, D.~Grauer, Y.~S.
  Hor, R.~J. Cava, M.~Z. Hasan, Observation of a large-gap
  topological-insulator class with a single dirac cone on the surface, Nat.\
  Phys. 5 (2009) 398.

\bibitem{Hsieh_Bi2Te3_Sb2Te3_PRL09}
D.~Hsieh, Y.~Xia, D.~Qian, L.~Wray, J.~H. Dil, F.~Meier, J.~Osterwalder,
  L.~Patthey, A.~V. Fedorov, H.~Lin, A.~Bansil, D.~Grauer, Y.~S. Hor, R.~J.
  Cava, M.~Z. Hasan, Time-reversal-protected single-dirac-cone
  topological-insulator states in bi2te3 and sb2te3: Topologically
  spin-polarized dirac fermions with pi berry's phase, Phys.\ Rev.\ Lett. 103
  (2009) 146401.

\bibitem{Taskin_TI_eh_PRL11}
A.~A. Taskin, Z.~Ren, S.~Sasaki, K.~Segawa, Y.~Ando,
  \href{http://link.aps.org/doi/10.1103/PhysRevLett.107.016801}{Observation of
  dirac holes and electrons in a topological insulator}, Phys. Rev. Lett. 107
  (2011) 016801.
\newblock \href {http://dx.doi.org/10.1103/PhysRevLett.107.016801}
  {\path{doi:10.1103/PhysRevLett.107.016801}}.
\newline\urlprefix\url{http://link.aps.org/doi/10.1103/PhysRevLett.107.016801}

\bibitem{Zhang_Bi2Se3_Film_Epitaxy_APL09}
G.~Zhang, H.~Qin, J.~Teng, J.~Guo, Q.~Guo, X.~Dai, Z.~Fang, K.~Wu,
  Quintuple-layer epitaxy of thin films of topological insulator bi2se3, Appl.\
  Phys.\ Lett. 95 (2009) 053114.

\bibitem{Wang_Bi2Te3_Ctrl_AM11}
G.~Wang, X.-G. Zhu, Y.-Y. Sun, Y.-Y. Li, T.~Zhang, J.~Wen, X.~Chen, K.~He,
  L.-L. Wang, X.-C. Ma, J.-F. Jia, S.~B. Zhang, Q.-K. Xue,
  \href{http://dx.doi.org/10.1002/adma.201100678}{Topological insulator thin
  films of bi2te3 with controlled electronic structure}, Advanced Materials
  23~(26) (2011) 2929.
\newblock \href {http://dx.doi.org/10.1002/adma.201100678}
  {\path{doi:10.1002/adma.201100678}}.
\newline\urlprefix\url{http://dx.doi.org/10.1002/adma.201100678}

\bibitem{Ren-PRB-2010}
Z.~Ren, A.~A. Taskin, S.~Sasaki, K.~Segawa, Y.~Ando, Phys. Rev. B 82 (2010)
  241306.

\bibitem{Benjamin_nature_2011}
B.~Sac{\'e}p{\'e}, J.~B. Oostinga, J.~Li, A.~Ubaldini, N.~J.~G. Couto,
  E.~Giannini, A.~F. Morpurgo,
  \href{http://dx.doi.org/10.1038/ncomms1586}{Gate-tuned normal and
  superconducting transport at the surface of a topological insulator}, Nat.
  Commun. 2 (2011) 575.
\newline\urlprefix\url{http://dx.doi.org/10.1038/ncomms1586}

\bibitem{Dohun_Surface_nphy_2012}
D.~Kim, S.~Cho, N.~P. Butch, P.~Syers, K.~Kirshenbaum, S.~Adam, J.~Paglione,
  M.~S. Fuhrer, \href{http://dx.doi.org/10.1038/nphys2286}{Surface conduction
  of topological dirac electrons in bulk insulating bi2se3}, Nat. Phys. 8
  (2012) 459.
\newline\urlprefix\url{http://dx.doi.org/10.1038/nphys2286}

\bibitem{Fuhrer_nature_2013}
D.~Kim, P.~Syers, N.~P. Butch, J.~Paglione, M.~S. Fuhrer,
  \href{http://dx.doi.org/10.1038/ncomms3040}{Coherent topological transport on
  the surface of bi2se3}, Nat. Commun. 4.
\newline\urlprefix\url{http://dx.doi.org/10.1038/ncomms3040}

\bibitem{Brune_prx_2014}
C.~Br{\"u}ne, C.~Thienel, M.~Stuiber, J.~B\"ottcher, H.~Buhmann, E.~G. Novik,
  C.-X. Liu, E.~M. Hankiewicz, L.~W. Molenkamp,
  \href{http://link.aps.org/doi/10.1103/PhysRevX.4.041045}{Dirac-screening
  stabilized surface-state transport in a topological insulator}, Phys. Rev. X
  4 (2014) 041045.
\newblock \href {http://dx.doi.org/10.1103/PhysRevX.4.041045}
  {\path{doi:10.1103/PhysRevX.4.041045}}.
\newline\urlprefix\url{http://link.aps.org/doi/10.1103/PhysRevX.4.041045}

\bibitem{Hellerstedt_APL_2014}
J.~Hellerstedt, M.~T. Edmonds, J.~H. Chen, W.~G. Cullen, C.~X. Zheng, M.~S.
  Fuhrer,
  \href{http://scitation.aip.org/content/aip/journal/apl/105/17/10.1063/1.4900749}{Thickness
  and growth-condition dependence of in-situ mobility and carrier density of
  epitaxial thin-film bi2se3}, Applied Physics Letters 105~(17) (2014) 173506.
\newblock \href {http://dx.doi.org/http://dx.doi.org/10.1063/1.4900749}
  {\path{doi:http://dx.doi.org/10.1063/1.4900749}}.
\newline\urlprefix\url{http://scitation.aip.org/content/aip/journal/apl/105/17/10.1063/1.4900749}

\bibitem{Lucas_nl_tran_2014}
L.~Barreto, L.~Khnemund, F.~Edler, C.~Tegenkamp, J.~Mi, M.~Bremholm, B.~B.
  Iversen, C.~Frydendahl, M.~Bianchi, P.~Hofmann,
  \href{http://dx.doi.org/10.1021/nl501489m}{Surface-dominated transport on a
  bulk topological insulator}, Nano Letters 14~(7) (2014) 3755.
\newblock \href {http://dx.doi.org/10.1021/nl501489m}
  {\path{doi:10.1021/nl501489m}}.
\newline\urlprefix\url{http://dx.doi.org/10.1021/nl501489m}

\bibitem{Burkov_TI_SpinCharge_PRL10}
A.~A. Burkov, D.~G. Hawthorn, Spin and charge transport on the surface of a
  topological insulator, Phys.\ Rev.\ Lett. 105 (2010) 066802.

\bibitem{Culcer_TI_Kineq_PRB10}
D.~Culcer, E.~H. Hwang, T.~D. Stanescu, S.~{Das Sarma}, Two-dimensional surface
  charge transport in topological insulators, Phys.\ Rev.\ B 82 (2010) 155457.

\bibitem{LiC.H_spin_polarization_nature2014}
C.~H. Li, O.~M.~J. van {`t}~Erve, J.~T. Robinson, Y.~Liu, L.~Li, B.~T. Jonker,
  \href{http://dx.doi.org/10.1038/nnano.2014.16}{Electrical detection of
  charge-current-induced spin polarization due to spin-momentum locking in
  bi2se3}, Nat. Nano. 9 (2014) 218.
\newline\urlprefix\url{http://dx.doi.org/10.1038/nnano.2014.16}

\bibitem{Jianshi_nl_spin_polarized_2014}
J.~Tang, L.-T. Chang, X.~Kou, K.~Murata, E.~S. Choi, M.~Lang, Y.~Fan, Y.~Jiang,
  M.~Montazeri, W.~Jiang, Y.~Wang, L.~He, K.~L. Wang,
  \href{http://dx.doi.org/10.1021/nl5026198}{Electrical detection of
  spin-polarized surface states conduction in (bi0.53sb0.47)2te3 topological
  insulator}, Nano Letters 14~(9) (2014) 5423.
\newblock \href {http://dx.doi.org/10.1021/nl5026198}
  {\path{doi:10.1021/nl5026198}}.
\newline\urlprefix\url{http://dx.doi.org/10.1021/nl5026198}

\bibitem{Kastl_surface_tran_nature_2015}
C.~Kastl, C.~Karnetzky, H.~Karl, A.~W. Holleitner,
  \href{http://dx.doi.org/10.1038/ncomms7617}{Ultrafast helicity control of
  surface currents in topological insulators with near-unity fidelity}, Nat.
  Commun. 6.
\newline\urlprefix\url{http://dx.doi.org/10.1038/ncomms7617}

\bibitem{Hor_DopedTI_FM_PRB10}
Y.~S. Hor, P.~Roushan, H.~Beidenkopf, J.~Seo, D.~Qu, J.~G. Checkelsky, L.~A.
  Wray, Y.~Xia, S.-Y. Xu, D.~Qian, M.~Z. Hasan, N.~P. Ong, A.~Yazdani, R.~J.
  Cava, The development of ferromagnetism in the doped topological insulator
  bi2-xmnxte3, Phys.\ Rev.\ B 81 (2010) 195203.

\bibitem{Jinsong-Zhang-Yayu-Wang-2013-science}
J.~Zhang, C.-Z. Chang, P.~Tang, Z.~Zhang, X.~Feng, K.~Li, L.-l. Wang, X.~Chen,
  C.~Liu, W.~Duan, K.~He, Q.-K. Xue, X.~Ma, Y.~Wang, Topology-driven magnetic
  quantum phase transition in topological insulators, Science 339~(6127) (2013)
  1582--1586.
\newblock \href {http://dx.doi.org/10.1126/science.1230905}
  {\path{doi:10.1126/science.1230905}}.

\bibitem{Collins-McIntyre_Cr_Bi2Se3_2014}
L.~J. Collins-McIntyre, S.~E. Harrison, P.~Sch{\"o}nherr, N.-J. Steinke, C.~J.
  Kinane, T.~R. Charlton, D.~Alba-Veneroa, A.~Pushp, A.~J. Kellock, S.~S.~P.
  Parkin, J.~S. Harris, S.~Langridge, G.~van~der Laan, T.~Hesjedal,
  \href{http://stacks.iop.org/0295-5075/107/i=5/a=57009}{Magnetic ordering in
  cr-doped bi 2 se 3 thin films}, EPL (Europhysics Letters) 107~(5) (2014)
  57009.
\newline\urlprefix\url{http://stacks.iop.org/0295-5075/107/i=5/a=57009}

\bibitem{Culcer_TI_AHE_PRB11}
D.~Culcer, S.~Das~Sarma,
  \href{http://link.aps.org/doi/10.1103/PhysRevB.83.245441}{Anomalous hall
  response of topological insulators}, Phys. Rev. B 83 (2011) 245441.
\newblock \href {http://dx.doi.org/10.1103/PhysRevB.83.245441}
  {\path{doi:10.1103/PhysRevB.83.245441}}.
\newline\urlprefix\url{http://link.aps.org/doi/10.1103/PhysRevB.83.245441}

\bibitem{Yu_TI_QuantAHE_Science10}
R.~Yu, W.~Zhang, H.~J. Zhang, S.~C. Zhang, X.~Dai, Z.~Fang, Quantized anomalous
  hall effect in magnetic topological insulators, Science 329 (2010) 61.

\bibitem{LuZhao_TITF_transport_PRB2013}
H.-Z. Lu, A.~Zhao, S.-Q. Shen,
  \href{http://link.aps.org/doi/10.1103/PhysRevLett.111.146802}{Quantum
  transport in magnetic topological insulator thin films}, Phys. Rev. Lett. 111
  (2013) 146802.
\newblock \href {http://dx.doi.org/10.1103/PhysRevLett.111.146802}
  {\path{doi:10.1103/PhysRevLett.111.146802}}.
\newline\urlprefix\url{http://link.aps.org/doi/10.1103/PhysRevLett.111.146802}

\bibitem{JiangQiao_TIF_QAHE_PRB2012}
H.~Jiang, Z.~Qiao, H.~Liu, Q.~Niu, Quantum anomalous hall effect with tunable
  chern number in magnetic topological insulator film, Phys. Rev. B 85 (2012)
  045445.
\newblock \href {http://dx.doi.org/10.1103/PhysRevB.85.045445}
  {\path{doi:10.1103/PhysRevB.85.045445}}.

\bibitem{Chang_TIF_ferromagnetism_AHE_AM2013}
C.-Z. Chang, J.~Zhang, M.~Liu, Z.~Zhang, X.~Feng, K.~Li, L.-L. Wang, X.~Chen,
  X.~Dai, Z.~Fang, X.-L. Qi, S.-C. Zhang, Y.~Wang, K.~He, X.-C. Ma, Q.-K. Xue,
  \href{http://dx.doi.org/10.1002/adma.201203493}{Thin films of magnetically
  doped topological insulator with carrier-independent long-range ferromagnetic
  order}, Advanced Materials 25~(7) (2013) 1065.
\newblock \href {http://dx.doi.org/10.1002/adma.201203493}
  {\path{doi:10.1002/adma.201203493}}.
\newline\urlprefix\url{http://dx.doi.org/10.1002/adma.201203493}

\bibitem{Chang_QAHE_exper_Science2013}
C.-Z. Chang, J.~Zhang, X.~Feng, J.~Shen, Z.~Zhang, M.~Guo, K.~Li, Y.~Ou,
  P.~Wei, L.-L. Wang, Z.-Q. Ji, Y.~Feng, S.~Ji, X.~Chen, J.~Jia, X.~Dai,
  Z.~Fang, S.-C. Zhang, K.~He, Y.~Wang, L.~Lu, X.-C. Ma, Q.-K. Xue,
  Experimental observation of the quantum anomalous hall effect in a magnetic
  topological insulator, Science 340 (2013) 167.

\bibitem{Sochnikov_TISC_nano_2013}
I.~Sochnikov, A.~J. Bestwick, J.~R. Williams, T.~M. Lippman, I.~R. Fisher,
  D.~Goldhaber-Gordon, J.~R. Kirtley, K.~A. Moler,
  \href{http://dx.doi.org/10.1021/nl400997k}{Direct measurement of
  current-phase relations in superconductor/topological
  insulator/superconductor junctions}, Nano Lett. 13~(7) (2013) 3086.
\newblock \href {http://dx.doi.org/10.1021/nl400997k}
  {\path{doi:10.1021/nl400997k}}.
\newline\urlprefix\url{http://dx.doi.org/10.1021/nl400997k}

\bibitem{Jin-Feng_TISC_2014}
M.-X. Wang, P.~Li, J.-P. Xu, Z.-L. Liu, J.-F. Ge, G.-Y. Wang, X.~Yang, Z.-A.
  Xu, S.-H. Ji, C.~L. Gao, D.~Qian, W.~Luo, C.~Liu, J.-F. Jia,
  \href{http://stacks.iop.org/1367-2630/16/i=12/a=123043}{Interface structure
  of a topological insulator/superconductor heterostructure}, New Journal of
  Physics 16~(12) (2014) 123043.
\newline\urlprefix\url{http://stacks.iop.org/1367-2630/16/i=12/a=123043}

\bibitem{SDS_TQC_RMP08}
C.~Nayak, S.~H. Simon, A.~Stern, M.~Freedman, S.~{Das Sarma}, Non-abelian
  anyons and topological quantum computation, Rev.\ Mod.\ Phys. 80 (2008) 1083.

\bibitem{Qiang-Hua_TISC_2014_SciRep}
Z.-Z. Li, F.-C. Zhang, Q.-H. Wang,
  \href{http://dx.doi.org/10.1038/srep06363}{Majorana modes in a topological
  insulator/s-wave superconductor heterostructure}, Sci. Rep. 4.
\newline\urlprefix\url{http://dx.doi.org/10.1038/srep06363}

\bibitem{Adam_2D_Tran_prb_2012}
S.~Adam, E.~H. Hwang, S.~Das~Sarma,
  \href{http://link.aps.org/doi/10.1103/PhysRevB.85.235413}{Two-dimensional
  transport and screening in topological insulator surface states}, Phys. Rev.
  B 85 (2012) 235413.
\newblock \href {http://dx.doi.org/10.1103/PhysRevB.85.235413}
  {\path{doi:10.1103/PhysRevB.85.235413}}.
\newline\urlprefix\url{http://link.aps.org/doi/10.1103/PhysRevB.85.235413}

\bibitem{Qiuzi_2Dtran_phimu_2012_prb}
Q.~Li, E.~Rossi, S.~Das~Sarma,
  \href{http://link.aps.org/doi/10.1103/PhysRevB.86.235443}{Two-dimensional
  electronic transport on the surface of three-dimensional topological
  insulators}, Phys. Rev. B 86 (2012) 235443.
\newblock \href {http://dx.doi.org/10.1103/PhysRevB.86.235443}
  {\path{doi:10.1103/PhysRevB.86.235443}}.
\newline\urlprefix\url{http://link.aps.org/doi/10.1103/PhysRevB.86.235443}

\bibitem{JMShao-oscillation}
J.~M. Shao, H.~Li, G.~W. Yang,
  \href{http://stacks.iop.org/0953-8984/25/i=42/a=425603}{Conductivity
  oscillation of surface state of three-dimensional topological insulators
  induced by a linearly polarized terahertz field}, Journal of Physics:
  Condensed Matter 25~(42) (2013) 425603.
\newline\urlprefix\url{http://stacks.iop.org/0953-8984/25/i=42/a=425603}

\bibitem{Costache_prl_2014}
M.~V. Costache, I.~Neumann, J.~F. Sierra, V.~Marinova, M.~M. Gospodinov,
  S.~Roche, S.~O. Valenzuela,
  \href{http://link.aps.org/doi/10.1103/PhysRevLett.112.086601}{Fingerprints of
  inelastic transport at the surface of the topological insulator
  ${\mathrm{bi}}_{2}{\mathrm{se}}_{3}$: Role of electron-phonon coupling},
  Phys. Rev. Lett. 112 (2014) 086601.
\newblock \href {http://dx.doi.org/10.1103/PhysRevLett.112.086601}
  {\path{doi:10.1103/PhysRevLett.112.086601}}.
\newline\urlprefix\url{http://link.aps.org/doi/10.1103/PhysRevLett.112.086601}

\bibitem{Kozlov_prl_2014}
D.~A. Kozlov, Z.~D. Kvon, E.~B. Olshanetsky, N.~N. Mikhailov, S.~A. Dvoretsky,
  D.~Weiss,
  \href{http://link.aps.org/doi/10.1103/PhysRevLett.112.196801}{Transport
  properties of a 3d topological insulator based on a strained high-mobility
  hgte film}, Phys. Rev. Lett. 112 (2014) 196801.
\newblock \href {http://dx.doi.org/10.1103/PhysRevLett.112.196801}
  {\path{doi:10.1103/PhysRevLett.112.196801}}.
\newline\urlprefix\url{http://link.aps.org/doi/10.1103/PhysRevLett.112.196801}

\bibitem{Syers_SmB_prl_2015}
P.~Syers, D.~Kim, M.~S. Fuhrer, J.~Paglione,
  \href{http://link.aps.org/doi/10.1103/PhysRevLett.114.096601}{Tuning bulk and
  surface conduction in the proposed topological kondo insulator
  ${\mathrm{smb}}_{6}$}, Phys. Rev. Lett. 114 (2015) 096601.
\newblock \href {http://dx.doi.org/10.1103/PhysRevLett.114.096601}
  {\path{doi:10.1103/PhysRevLett.114.096601}}.
\newline\urlprefix\url{http://link.aps.org/doi/10.1103/PhysRevLett.114.096601}

\bibitem{Kim-Hall2013}
D.~J. Kim, S.~Thomas, T.~Grant, J.~Botimer, Z.~Fisk, J.~Xia,
  \href{http://dx.doi.org/10.1038/srep03150}{Surface hall effect and nonlocal
  transport in smb6: Evidence for surface conduction}, Sci. Rep. 3.
\newline\urlprefix\url{http://dx.doi.org/10.1038/srep03150}

\bibitem{Jing_Shoucheng_QAHE_2014_prb}
J.~Wang, B.~Lian, S.-C. Zhang,
  \href{http://link.aps.org/doi/10.1103/PhysRevB.89.085106}{Universal scaling
  of the quantum anomalous hall plateau transition}, Phys. Rev. B 89 (2014)
  085106.
\newblock \href {http://dx.doi.org/10.1103/PhysRevB.89.085106}
  {\path{doi:10.1103/PhysRevB.89.085106}}.
\newline\urlprefix\url{http://link.aps.org/doi/10.1103/PhysRevB.89.085106}

\bibitem{ShiJunren_QAHE_prl_2014}
Y.~Zhang, J.~Shi,
  \href{http://link.aps.org/doi/10.1103/PhysRevLett.113.016801}{Quantum
  anomalous hall insulator of composite fermions}, Phys. Rev. Lett. 113 (2014)
  016801.
\newblock \href {http://dx.doi.org/10.1103/PhysRevLett.113.016801}
  {\path{doi:10.1103/PhysRevLett.113.016801}}.
\newline\urlprefix\url{http://link.aps.org/doi/10.1103/PhysRevLett.113.016801}

\bibitem{Dohun_Thermoelectric_nl_2014}
D.~Kim, P.~Syers, N.~P. Butch, J.~Paglione, M.~S. Fuhrer,
  \href{http://dx.doi.org/10.1021/nl4032154}{Ambipolar surface state
  thermoelectric power of topological insulator bi2se3}, Nano Letters 14~(4)
  (2014) 1701.
\newblock \href {http://dx.doi.org/10.1021/nl4032154}
  {\path{doi:10.1021/nl4032154}}.
\newline\urlprefix\url{http://dx.doi.org/10.1021/nl4032154}

\bibitem{Durst_2015}
A.~C. Durst,
  \href{http://link.aps.org/doi/10.1103/PhysRevB.91.094519}{Low-temperature
  thermal transport at the interface of a topological insulator and a $d$-wave
  superconductor}, Phys. Rev. B 91 (2015) 094519.
\newblock \href {http://dx.doi.org/10.1103/PhysRevB.91.094519}
  {\path{doi:10.1103/PhysRevB.91.094519}}.
\newline\urlprefix\url{http://link.aps.org/doi/10.1103/PhysRevB.91.094519}

\bibitem{ZhangJinsong_prb_2015}
J.~Zhang, X.~Feng, Y.~Xu, M.~Guo, Z.~Zhang, Y.~Ou, Y.~Feng, K.~Li, H.~Zhang,
  L.~Wang, X.~Chen, Z.~Gan, S.-C. Zhang, K.~He, X.~Ma, Q.-K. Xue, Y.~Wang,
  \href{http://link.aps.org/doi/10.1103/PhysRevB.91.075431}{Disentangling the
  magnetoelectric and thermoelectric transport in topological insulator thin
  films}, Phys. Rev. B 91 (2015) 075431.
\newblock \href {http://dx.doi.org/10.1103/PhysRevB.91.075431}
  {\path{doi:10.1103/PhysRevB.91.075431}}.
\newline\urlprefix\url{http://link.aps.org/doi/10.1103/PhysRevB.91.075431}

\bibitem{He_Bi2Te3_Film_WAL_ImpEff_PRL11}
H.-T. He, G.~Wang, T.~Zhang, I.-K. Sou, G.~K.~L. Wong, J.-N. Wang, H.-Z. Lu,
  S.-Q. Shen, F.-C. Zhang,
  \href{http://link.aps.org/doi/10.1103/PhysRevLett.106.166805}{Impurity effect
  on weak antilocalization in the topological insulator
  ${\mathrm{bi}}_{2}{\mathrm{te}}_{3}$}, Phys. Rev. Lett. 106 (2011) 166805.
\newblock \href {http://dx.doi.org/10.1103/PhysRevLett.106.166805}
  {\path{doi:10.1103/PhysRevLett.106.166805}}.
\newline\urlprefix\url{http://link.aps.org/doi/10.1103/PhysRevLett.106.166805}

\bibitem{Cheng_TI_STM_LL_PRL10}
P.~Cheng, C.~Song, T.~Zhang, Y.~Zhang, Y.~Wang, J.-F. Jia, J.~Wang, Y.~Wang,
  B.-F. Zhu, X.~Chen, X.~Ma, K.~He, L.~Wang, X.~Dai, Z.~Fang, X.~Xie, X.-L. Qi,
  C.-X. Liu, S.-C. Zhang, Q.-K. Xue, Landau quantization of massless dirac
  fermions in topological insulator, Phys.\ Rev.\ Lett. 105 (2010) 076801.

\bibitem{Tkachov_HgTe_WAL_PRB11}
G.~Tkachov, E.~M. Hankiewicz, Weak antilocalization in hgte quantum wells and
  topological surface states: Massive versus massless dirac fermions, Phys.\
  Rev.\ B 84 (2011) 035444.

\bibitem{Competition_WL_WAL_2011}
H.-Z. Lu, J.~Shi, S.-Q. Shen,
  \href{http://link.aps.org/doi/10.1103/PhysRevLett.107.076801}{Competition
  between weak localization and antilocalization in topological surface
  states}, Phys. Rev. Lett. 107 (2011) 076801.
\newblock \href {http://dx.doi.org/10.1103/PhysRevLett.107.076801}
  {\path{doi:10.1103/PhysRevLett.107.076801}}.
\newline\urlprefix\url{http://link.aps.org/doi/10.1103/PhysRevLett.107.076801}

\bibitem{Culcer_TI_Int_PRB11}
D.~Culcer, \href{http://link.aps.org/doi/10.1103/PhysRevB.84.235411}{Linear
  response theory of interacting topological insulators}, Phys. Rev. B 84
  (2011) 235411.
\newblock \href {http://dx.doi.org/10.1103/PhysRevB.84.235411}
  {\path{doi:10.1103/PhysRevB.84.235411}}.
\newline\urlprefix\url{http://link.aps.org/doi/10.1103/PhysRevB.84.235411}

\bibitem{Yamaji-PRB-2011}
Y.~Yamaji, M.~Imada, Mott physics on helical edges of two-dimensional
  topological insulators, Phys. Rev. B 83 (2011) 205122.

\bibitem{Peters-PRL-2012}
R.~Peters, N.~Kawakami, T.~Pruschke, Spin-selective kondo insulator:
  Cooperation of ferromagnetism and the kondo effect, Phys. Rev. Lett. 108
  (2012) 086402.

\bibitem{Yoshida-PRB-2012}
T.~Yoshida, S.~Fujimoto, N.~Kawakami,
  \href{http://link.aps.org/doi/10.1103/PhysRevB.85.125113}{Correlation effects
  on a topological insulator at finite temperatures}, Phys. Rev. B 85 (2012)
  125113.
\newblock \href {http://dx.doi.org/10.1103/PhysRevB.85.125113}
  {\path{doi:10.1103/PhysRevB.85.125113}}.
\newline\urlprefix\url{http://link.aps.org/doi/10.1103/PhysRevB.85.125113}

\bibitem{Ostrovsky_TI_IntCrit_PRL10}
P.~M. Ostrovsky, I.~V. Gornyi, A.~D. Mirlin,
  \href{http://link.aps.org/doi/10.1103/PhysRevLett.105.036803}{Interaction-induced
  criticality in ${\mathbb{z}}_{2}$ topological insulators}, Phys. Rev. Lett.
  105 (2010) 036803.
\newblock \href {http://dx.doi.org/10.1103/PhysRevLett.105.036803}
  {\path{doi:10.1103/PhysRevLett.105.036803}}.
\newline\urlprefix\url{http://link.aps.org/doi/10.1103/PhysRevLett.105.036803}

\bibitem{WangCulcer_TI_Kondo_PRB2013}
J.~Wang, D.~Culcer, Suppression of the kondo resistivity minimum in topological
  insulators, Phys. Rev. B 88 (2013) 125140.
\newblock \href {http://dx.doi.org/10.1103/PhysRevB.88.125140}
  {\path{doi:10.1103/PhysRevB.88.125140}}.

\bibitem{LiQiuzi_2013_prb}
S.~Das~Sarma, Q.~Li,
  \href{http://link.aps.org/doi/10.1103/PhysRevB.88.081404}{Many-body effects
  and possible superconductivity in the two-dimensional metallic surface states
  of three-dimensional topological insulators}, Phys. Rev. B 88 (2013) 081404.
\newblock \href {http://dx.doi.org/10.1103/PhysRevB.88.081404}
  {\path{doi:10.1103/PhysRevB.88.081404}}.
\newline\urlprefix\url{http://link.aps.org/doi/10.1103/PhysRevB.88.081404}

\bibitem{Hai-Zhou_Conductivity_2014_prl}
H.-Z. Lu, S.-Q. Shen,
  \href{http://link.aps.org/doi/10.1103/PhysRevLett.112.146601}{Finite-temperature
  conductivity and magnetoconductivity of topological insulators}, Phys. Rev.
  Lett. 112 (2014) 146601.
\newblock \href {http://dx.doi.org/10.1103/PhysRevLett.112.146601}
  {\path{doi:10.1103/PhysRevLett.112.146601}}.
\newline\urlprefix\url{http://link.aps.org/doi/10.1103/PhysRevLett.112.146601}

\bibitem{1977_drag}
M.~Pograbinskii, Mutual drag of carriers in a
  semiconductor-insulator-semiconductor system, Sov.Phys.Semicond. 11 (1977)
  372.

\bibitem{1995_Hall_drag}
A.~Kamenev, Y.~Oreg,
  \href{http://link.aps.org/doi/10.1103/PhysRevB.52.7516}{Coulomb drag in
  normal metals and superconductors: Diagrammatic approach}, Phys. Rev. B 52
  (1995) 7516.
\newblock \href {http://dx.doi.org/10.1103/PhysRevB.52.7516}
  {\path{doi:10.1103/PhysRevB.52.7516}}.
\newline\urlprefix\url{http://link.aps.org/doi/10.1103/PhysRevB.52.7516}

\bibitem{1999_Hall_drag}
A.~G. Rojo, \href{http://stacks.iop.org/0953-8984/11/i=5/a=004}{Electron-drag
  effects in coupled electron systems}, Journal of Physics: Condensed Matter
  11~(5) (1999) R31.
\newline\urlprefix\url{http://stacks.iop.org/0953-8984/11/i=5/a=004}

\bibitem{Tse_SO_Drag_PRB07}
W.-K. Tse, S.~Das~Sarma,
  \href{http://link.aps.org/doi/10.1103/PhysRevB.75.045333}{Coulomb drag and
  spin drag in the presence of spin-orbit coupling}, Phys. Rev. B 75 (2007)
  045333.
\newblock \href {http://dx.doi.org/10.1103/PhysRevB.75.045333}
  {\path{doi:10.1103/PhysRevB.75.045333}}.
\newline\urlprefix\url{http://link.aps.org/doi/10.1103/PhysRevB.75.045333}

\bibitem{Katsnelson2011}
M.~I. Katsnelson,
  \href{http://link.aps.org/doi/10.1103/PhysRevB.84.041407}{Coulomb drag in
  graphene single layers separated by a thin spacer}, Phys. Rev. B 84 (2011)
  041407.
\newblock \href {http://dx.doi.org/10.1103/PhysRevB.84.041407}
  {\path{doi:10.1103/PhysRevB.84.041407}}.
\newline\urlprefix\url{http://link.aps.org/doi/10.1103/PhysRevB.84.041407}

\bibitem{Hwang2011}
E.~H. Hwang, R.~Sensarma, S.~Das~Sarma,
  \href{http://link.aps.org/doi/10.1103/PhysRevB.84.245441}{Coulomb drag in
  monolayer and bilayer graphene}, Phys. Rev. B 84 (2011) 245441.
\newblock \href {http://dx.doi.org/10.1103/PhysRevB.84.245441}
  {\path{doi:10.1103/PhysRevB.84.245441}}.
\newline\urlprefix\url{http://link.aps.org/doi/10.1103/PhysRevB.84.245441}

\bibitem{Tse2007}
W.-K. Tse, B.~Y.-K. Hu, S.~Das~Sarma,
  \href{http://link.aps.org/doi/10.1103/PhysRevB.76.081401}{Theory of coulomb
  drag in graphene}, Phys. Rev. B 76 (2007) 081401.
\newblock \href {http://dx.doi.org/10.1103/PhysRevB.76.081401}
  {\path{doi:10.1103/PhysRevB.76.081401}}.
\newline\urlprefix\url{http://link.aps.org/doi/10.1103/PhysRevB.76.081401}

\bibitem{Amorim_drag}
B.~Amorim, N.~M.~R. Peres,
  \href{http://stacks.iop.org/0953-8984/24/i=33/a=335602}{On coulomb drag in
  double layer systems}, Journal of Physics: Condensed Matter 24~(33) (2012)
  335602.
\newline\urlprefix\url{http://stacks.iop.org/0953-8984/24/i=33/a=335602}

\bibitem{M.Carrega2012}
M.~Carrega, T.~Tudorovskiy, A.~Principi, M.~I. Katsnelson, M.~Polini,
  \href{http://stacks.iop.org/1367-2630/14/i=6/a=063033}{Theory of coulomb drag
  for massless dirac fermions}, New Journal of Physics 14~(6) (2012) 063033.
\newline\urlprefix\url{http://stacks.iop.org/1367-2630/14/i=6/a=063033}

\bibitem{Kim2011}
S.~Kim, I.~Jo, J.~Nah, Z.~Yao, S.~K. Banerjee, E.~Tutuc,
  \href{http://link.aps.org/doi/10.1103/PhysRevB.83.161401}{Coulomb drag of
  massless fermions in graphene}, Phys. Rev. B 83 (2011) 161401.
\newblock \href {http://dx.doi.org/10.1103/PhysRevB.83.161401}
  {\path{doi:10.1103/PhysRevB.83.161401}}.
\newline\urlprefix\url{http://link.aps.org/doi/10.1103/PhysRevB.83.161401}

\bibitem{Narozhny_drag_2012}
B.~N. Narozhny, M.~Titov, I.~V. Gornyi, P.~M. Ostrovsky,
  \href{http://link.aps.org/doi/10.1103/PhysRevB.85.195421}{Coulomb drag in
  graphene: Perturbation theory}, Phys. Rev. B 85 (2012) 195421.
\newblock \href {http://dx.doi.org/10.1103/PhysRevB.85.195421}
  {\path{doi:10.1103/PhysRevB.85.195421}}.
\newline\urlprefix\url{http://link.aps.org/doi/10.1103/PhysRevB.85.195421}

\bibitem{Kim2012}
S.~Kim, E.~Tutuc,
  \href{http://www.sciencedirect.com/science/article/pii/S0038109812002384}{Coulomb
  drag and magnetotransport in graphene double layers}, Solid State
  Communications 152~(15) (2012) 1283.
\newblock \href {http://dx.doi.org/http://dx.doi.org/10.1016/j.ssc.2012.04.032}
  {\path{doi:http://dx.doi.org/10.1016/j.ssc.2012.04.032}}.
\newline\urlprefix\url{http://www.sciencedirect.com/science/article/pii/S0038109812002384}

\bibitem{Gorbachevi_nature_2012}
R.~V. Gorbachev, A.~K. Geim, M.~I. Katsnelson, K.~S. Novoselov, T.~Tudorovskiy,
  I.~V. Grigorieva, A.~H. MacDonald, S.~V. Morozov, K.~Watanabe, T.~Taniguchi,
  L.~A. Ponomarenko, \href{http://dx.doi.org/10.1038/nphys2441}{Strong coulomb
  drag and broken symmetry in double-layer graphene}, Nat. Phys. 8 (2012) 896.
\newline\urlprefix\url{http://dx.doi.org/10.1038/nphys2441}

\bibitem{Sch_drag_prl_2013}
M.~Sch{\"u}tt, P.~M. Ostrovsky, M.~Titov, I.~V. Gornyi, B.~N. Narozhny, A.~D.
  Mirlin, \href{http://link.aps.org/doi/10.1103/PhysRevLett.110.026601}{Coulomb
  drag in graphene near the dirac point}, Phys. Rev. Lett. 110 (2013) 026601.
\newblock \href {http://dx.doi.org/10.1103/PhysRevLett.110.026601}
  {\path{doi:10.1103/PhysRevLett.110.026601}}.
\newline\urlprefix\url{http://link.aps.org/doi/10.1103/PhysRevLett.110.026601}

\bibitem{Gamucci_nature_2014}
A.~Gamucci, D.~Spirito, M.~Carrega, B.~Karmakar, A.~Lombardo, M.~Bruna, L.~N.
  Pfeiffer, K.~W. West, A.~C. Ferrari, M.~Polini, V.~Pellegrini,
  \href{http://dx.doi.org/10.1038/ncomms6824}{Anomalous low-temperature coulomb
  drag in graphene-gaas heterostructures}, Nat. Commun. 5.
\newline\urlprefix\url{http://dx.doi.org/10.1038/ncomms6824}

\bibitem{Hall_drag_Graphene}
M.~Titov, R.~V. Gorbachev, B.~N. Narozhny, T.~Tudorovskiy, M.~Sch{\"u}tt, P.~M.
  Ostrovsky, I.~V. Gornyi, A.~D. Mirlin, M.~I. Katsnelson, K.~S. Novoselov,
  A.~K. Geim, L.~A. Ponomarenko,
  \href{http://link.aps.org/doi/10.1103/PhysRevLett.111.166601}{Giant
  magnetodrag in graphene at charge neutrality}, Phys. Rev. Lett. 111 (2013)
  166601.
\newblock \href {http://dx.doi.org/10.1103/PhysRevLett.111.166601}
  {\path{doi:10.1103/PhysRevLett.111.166601}}.
\newline\urlprefix\url{http://link.aps.org/doi/10.1103/PhysRevLett.111.166601}

\bibitem{Song_Halldrag_2013}
J.~C.~W. Song, L.~S. Levitov,
  \href{http://link.aps.org/doi/10.1103/PhysRevLett.111.126601}{Hall drag and
  magnetodrag in graphene}, Phys. Rev. Lett. 111 (2013) 126601.
\newblock \href {http://dx.doi.org/10.1103/PhysRevLett.111.126601}
  {\path{doi:10.1103/PhysRevLett.111.126601}}.
\newline\urlprefix\url{http://link.aps.org/doi/10.1103/PhysRevLett.111.126601}

\bibitem{Vasko}
F.~T. Vasko, O.~E. Raichev, Quantum Kinetic Theory and Applications, Springer,
  New York, 2005.

\bibitem{MWWu_Peng}
P.Zhang, M.W.Wu, Phys. Rev. B 87 (2013) 085319.

\bibitem{SDS_Gfn_RMP11}
S.~{Das Sarma}, S.~Adam, E.~H. Hwang, E.~Rossi, Electronic transport in two
  dimensional graphene, Rev.\ Mod.\ Phys. 83 (2011) 407.

\bibitem{Weizhe_TITF_2014_prb}
W.~E. Liu, H.~Liu, D.~Culcer, Phys. Rev. B 89 (2014) 195417.

\bibitem{LuShan_TITF_MassiveDirac_spinphys_PRB2010}
H.-Z. Lu, W.-Y. Shan, W.~Yao, Q.~Niu, S.-Q. Shen, Massive dirac fermions and
  spin physics in an ultrathin film of topological insulator, Phys. Rev. B 81
  (2010) 115407.
\newblock \href {http://dx.doi.org/10.1103/PhysRevB.81.115407}
  {\path{doi:10.1103/PhysRevB.81.115407}}.

\bibitem{Polini_Exciton_Drag_PRL2012}
M.~P. Mink, H.~T.~C. Stoof, R.~A. Duine, M.~Polini, G.~Vignale,
  \href{http://link.aps.org/doi/10.1103/PhysRevLett.108.186402}{Probing the
  topological exciton condensate via coulomb drag}, Phys. Rev. Lett. 108 (2012)
  186402.
\newblock \href {http://dx.doi.org/10.1103/PhysRevLett.108.186402}
  {\path{doi:10.1103/PhysRevLett.108.186402}}.
\newline\urlprefix\url{http://link.aps.org/doi/10.1103/PhysRevLett.108.186402}

\bibitem{Pellegrini_Exciton_NCom2014}
A.~Gamucci, D.~Spirito, M.~Carrega, B.~Karmakar, A.~Lombardo, M.~Bruna, L.~N.
  Pfeiffer, K.~W. West, A.~C. Ferrari, M.~Polini, V.~Pellegrini,
  \href{http://dx.doi.org/10.1038/ncomms6824}{Anomalous low-temperature coulomb
  drag in graphene-gaas heterostructures}, Nat. Commun. 5 (2014) 5824.
\newline\urlprefix\url{http://dx.doi.org/10.1038/ncomms6824}

\bibitem{Beidenkopf_NP11}
H.~Beidenkopf, P.~Roushan, J.~Seo, L.~Gorman, I.~Drozdov, Y.~S. Hor, R.~J.
  Cava, A.~Yazdani, Spatial fluctuations of helical dirac fermions on the
  surface of topological insulators, Nat. Phys. 7 (2011) 939.

\bibitem{Kim_TI_e-ph_PRL12}
D.~Kim, Q.~L.~P. Syers, N.~P. Butch, J.~Paglione, S.~{Das Sarma}, M.~S. Fuhrer,
  Phys.\ Rev.\ Lett. 109 (2012) 166801.

\end{thebibliography}

\end{document}